\documentclass[twocolumn]{aastex631}
\usepackage{hyperref}
\usepackage{verbatim}
\usepackage{tabularx}

\usepackage{amsmath}
\usepackage{multirow}
\usepackage{graphicx}
\usepackage{amssymb}
\usepackage{bm}	
\usepackage{nccmath}
\usepackage[T1]{fontenc}
\usepackage{ae,aecompl}
\usepackage{newtxtext,newtxmath}
\usepackage{xcolor}
\usepackage{appendix}

\newcommand{\tb}[1]{\textcolor{black}{#1}}
\newcommand{\modify}[1]{\textcolor{black}{#1}}
\newcommand{\package}[1]{\texttt{#1}}
\newcommand{\target}[1]{\texttt{#1}}
\newcommand{\kms}{\,km\,s$^{-1}$}
\newcommand{\angstrom}{\mbox{\normalfont\AA\,}}
\newcommand{\bvls}{\textit{BVLS }}
\newcommand{\oii}{[\ion{O}{2}] }
\newcommand{\oiii}{[\ion{O}{3}] }

\shorttitle{Archetype-Based Redshift Estimation for DESI}
\shortauthors{Anand et al.}

\begin{document}

\title{Archetype-Based Redshift Estimation for the Dark Energy Spectroscopic Instrument Survey}

% Author list file generated with: mkauthlist 1.3.0+37.g17f199c 
% mkauthlist -f --sort-firsttier --orcid -j apj DESI-2024-0422_author_list.csv author_archetype_May7.tex 
%% Orcid numbers may need \usepackage{orcidlink}.
%% Use \input to call the file

\author[0000-0003-2923-1585]{Abhijeet~Anand}
\affiliation{Lawrence Berkeley National Laboratory, 1 Cyclotron Road, Berkeley, CA 94720, USA}

\author[0000-0001-9822-6793]{Julien~Guy}
\affiliation{Lawrence Berkeley National Laboratory, 1 Cyclotron Road, Berkeley, CA 94720, USA}

\author[0000-0003-4162-6619]{Stephen~Bailey}
\affiliation{Lawrence Berkeley National Laboratory, 1 Cyclotron Road, Berkeley, CA 94720, USA}

\author[0000-0002-2733-4559]{John~Moustakas}
\affiliation{Department of Physics and Astronomy, Siena College, 515 Loudon Road, Loudonville, NY 12211, USA}

\author{J.~Aguilar}
\affiliation{Lawrence Berkeley National Laboratory, 1 Cyclotron Road, Berkeley, CA 94720, USA}

\author[0000-0001-6098-7247]{S.~Ahlen}
\affiliation{Physics Dept., Boston University, 590 Commonwealth Avenue, Boston, MA 02215, USA}

\author[0000-0002-9836-603X]{A.~S.~Bolton}
\affiliation{NSF NOIRLab, 950 N. Cherry Ave., Tucson, AZ 85719, USA}

\author[0000-0002-8934-0954]{A.~Brodzeller}
\affiliation{Department of Physics and Astronomy, The University of Utah, 115 South 1400 East, Salt Lake City, UT 84112, USA}
\affiliation{Lawrence Berkeley National Laboratory, 1 Cyclotron Road, Berkeley, CA 94720, USA}

\author{D.~Brooks}
\affiliation{Department of Physics \& Astronomy, University College London, Gower Street, London, WC1E 6BT, UK}

\author{T.~Claybaugh}
\affiliation{Lawrence Berkeley National Laboratory, 1 Cyclotron Road, Berkeley, CA 94720, USA}

\author[0000-0002-5954-7903]{S.~Cole}
\affiliation{Institute for Computational Cosmology, Department of Physics, Durham University, South Road, Durham DH1 3LE, UK}

\author[0000-0002-1769-1640]{A.~de la Macorra}
\affiliation{Instituto de F\'{\i}sica, Universidad Nacional Aut\'{o}noma de M\'{e}xico,  Cd. de M\'{e}xico  C.P. 04510,  M\'{e}xico}

\author[0000-0002-5665-7912]{Biprateep~Dey}
\affiliation{Department of Physics \& Astronomy and Pittsburgh Particle Physics, Astrophysics, and Cosmology Center (PITT PACC), University of Pittsburgh, 3941 O'Hara Street, Pittsburgh, PA 15260, USA}

\author[0000-0003-2371-3356]{K.~Fanning}
\affiliation{Kavli Institute for Particle Astrophysics and Cosmology, Stanford University, Menlo Park, CA 94305, USA}
\affiliation{SLAC National Accelerator Laboratory, Menlo Park, CA 94305, USA}

\author[0000-0002-2890-3725]{J.~E.~Forero-Romero}
\affiliation{Departamento de F\'isica, Universidad de los Andes, Cra. 1 No. 18A-10, Edificio Ip, CP 111711, Bogot\'a, Colombia}
\affiliation{Observatorio Astron\'omico, Universidad de los Andes, Cra. 1 No. 18A-10, Edificio H, CP 111711 Bogot\'a, Colombia}

\author{E.~Gaztañaga}
\affiliation{Institut d'Estudis Espacials de Catalunya (IEEC), 08034 Barcelona, Spain}
\affiliation{Institute of Cosmology and Gravitation, University of Portsmouth, Dennis Sciama Building, Portsmouth, PO1 3FX, UK}
\affiliation{Institute of Space Sciences, ICE-CSIC, Campus UAB, Carrer de Can Magrans s/n, 08913 Bellaterra, Barcelona, Spain}

\author[0000-0003-3142-233X]{S.~Gontcho A Gontcho}
\affiliation{Lawrence Berkeley National Laboratory, 1 Cyclotron Road, Berkeley, CA 94720, USA}

\author{G.~Gutierrez}
\affiliation{Fermi National Accelerator Laboratory, PO Box 500, Batavia, IL 60510, USA}

\author{K.~Honscheid}
\affiliation{Center for Cosmology and AstroParticle Physics, The Ohio State University, 191 West Woodruff Avenue, Columbus, OH 43210, USA}
\affiliation{Department of Physics, The Ohio State University, 191 West Woodruff Avenue, Columbus, OH 43210, USA}
\affiliation{The Ohio State University, Columbus, 43210 OH, USA}

\author[0000-0002-1081-9410]{C.~Howlett}
\affiliation{School of Mathematics and Physics, University of Queensland, 4072, Australia}

\author{S.~Juneau}
\affiliation{NSF NOIRLab, 950 N. Cherry Ave., Tucson, AZ 85719, USA}

\author[0000-0002-8828-5463]{D.~Kirkby}
\affiliation{Department of Physics and Astronomy, University of California, Irvine, 92697, USA}

\author[0000-0003-3510-7134]{T.~Kisner}
\affiliation{Lawrence Berkeley National Laboratory, 1 Cyclotron Road, Berkeley, CA 94720, USA}

\author[0000-0001-6356-7424]{A.~Kremin}
\affiliation{Lawrence Berkeley National Laboratory, 1 Cyclotron Road, Berkeley, CA 94720, USA}

\author{A.~Lambert}
\affiliation{Lawrence Berkeley National Laboratory, 1 Cyclotron Road, Berkeley, CA 94720, USA}

\author[0000-0003-1838-8528]{M.~Landriau}
\affiliation{Lawrence Berkeley National Laboratory, 1 Cyclotron Road, Berkeley, CA 94720, USA}

\author[0000-0001-7178-8868]{L.~Le~Guillou}
\affiliation{Sorbonne Universit\'{e}, CNRS/IN2P3, Laboratoire de Physique Nucl\'{e}aire et de Hautes Energies (LPNHE), FR-75005 Paris, France}

\author[0000-0003-4962-8934]{M.~Manera}
\affiliation{Departament de F\'{i}sica, Serra H\'{u}nter, Universitat Aut\`{o}noma de Barcelona, 08193 Bellaterra (Barcelona), Spain}
\affiliation{Institut de F\'{i}sica d’Altes Energies (IFAE), The Barcelona Institute of Science and Technology, Campus UAB, 08193 Bellaterra Barcelona, Spain}

\author[0000-0002-1125-7384]{A.~Meisner}
\affiliation{NSF NOIRLab, 950 N. Cherry Ave., Tucson, AZ 85719, USA}

\author{R.~Miquel}
\affiliation{Instituci\'{o} Catalana de Recerca i Estudis Avan\c{c}ats, Passeig de Llu\'{\i}s Companys, 23, 08010 Barcelona, Spain}
\affiliation{Institut de F\'{i}sica d’Altes Energies (IFAE), The Barcelona Institute of Science and Technology, Campus UAB, 08193 Bellaterra Barcelona, Spain}

\author{E.~Mueller}
\affiliation{Department of Physics and Astronomy, University of Sussex, Brighton BN1 9QH, U.K}

\author[0000-0002-1544-8946]{G.~Niz}
\affiliation{Departamento de F\'{i}sica, Universidad de Guanajuato - DCI, C.P. 37150, Leon, Guanajuato, M\'{e}xico}
\affiliation{Instituto Avanzado de Cosmolog\'{\i}a A.~C., San Marcos 11 - Atenas 202. Magdalena Contreras, 10720. Ciudad de M\'{e}xico, M\'{e}xico}

\author[0000-0003-3188-784X]{N.~Palanque-Delabrouille}
\affiliation{IRFU, CEA, Universit\'{e} Paris-Saclay, F-91191 Gif-sur-Yvette, France}
\affiliation{Lawrence Berkeley National Laboratory, 1 Cyclotron Road, Berkeley, CA 94720, USA}

\author[0000-0002-0644-5727]{W.~J.~Percival}
\affiliation{Department of Physics and Astronomy, University of Waterloo, 200 University Ave W, Waterloo, ON N2L 3G1, Canada}
\affiliation{Perimeter Institute for Theoretical Physics, 31 Caroline St. North, Waterloo, ON N2L 2Y5, Canada}
\affiliation{Waterloo Centre for Astrophysics, University of Waterloo, 200 University Ave W, Waterloo, ON N2L 3G1, Canada}

\author{C.~Poppett}
\affiliation{Lawrence Berkeley National Laboratory, 1 Cyclotron Road, Berkeley, CA 94720, USA}
\affiliation{Space Sciences Laboratory, University of California, Berkeley, 7 Gauss Way, Berkeley, CA  94720, USA}
\affiliation{University of California, Berkeley, 110 Sproul Hall \#5800 Berkeley, CA 94720, USA}

\author[0000-0001-7145-8674]{F.~Prada}
\affiliation{Instituto de Astrof\'{i}sica de Andaluc\'{i}a (CSIC), Glorieta de la Astronom\'{i}a, s/n, E-18008 Granada, Spain}

\author[0000-0001-5999-7923]{A.~Raichoor}
\affiliation{Lawrence Berkeley National Laboratory, 1 Cyclotron Road, Berkeley, CA 94720, USA}

\author[0000-0001-5589-7116]{M.~Rezaie}
\affiliation{Department of Physics, Kansas State University, 116 Cardwell Hall, Manhattan, KS 66506, USA}

\author{G.~Rossi}
\affiliation{Department of Physics and Astronomy, Sejong University, Seoul, 143-747, Korea}

\author[0000-0002-9646-8198]{E.~Sanchez}
\affiliation{CIEMAT, Avenida Complutense 40, E-28040 Madrid, Spain}

\author[0000-0002-3569-7421]{E.~F.~Schlafly}
\affiliation{Space Telescope Science Institute, 3700 San Martin Drive, Baltimore, MD 21218, USA}

\author{D.~Schlegel}
\affiliation{Lawrence Berkeley National Laboratory, 1 Cyclotron Road, Berkeley, CA 94720, USA}

\author{M.~Schubnell}
\affiliation{Department of Physics, University of Michigan, Ann Arbor, MI 48109, USA}
\affiliation{University of Michigan, Ann Arbor, MI 48109, USA}

\author{D.~Sprayberry}
\affiliation{NSF NOIRLab, 950 N. Cherry Ave., Tucson, AZ 85719, USA}

\author[0000-0003-1704-0781]{G.~Tarl\'{e}}
\affiliation{University of Michigan, Ann Arbor, MI 48109, USA}

\author{C.~Warner}
\affiliation{Department of Astronomy, University of Florida, 211 Bryant Space Science Center, Gainesville, FL 32611, USA}

\author{B.~A.~Weaver}
\affiliation{NSF NOIRLab, 950 N. Cherry Ave., Tucson, AZ 85719, USA}

\author[0000-0001-5381-4372]{R.~Zhou}
\affiliation{Lawrence Berkeley National Laboratory, 1 Cyclotron Road, Berkeley, CA 94720, USA}

\author[0000-0002-6684-3997]{H.~Zou}
\affiliation{National Astronomical Observatories, Chinese Academy of Sciences, A20 Datun Rd., Chaoyang District, Beijing, 100012, P.R. China}

\correspondingauthor{Abhijeet~Anand}
\email{AbhijeetAnand@lbl.gov}

\begin{abstract}

We present a computationally efficient galaxy archetype-based redshift estimation and spectral classification method for the Dark Energy Survey Instrument (DESI) survey. The DESI survey currently relies on a redshift fitter and spectral classifier using a linear combination of PCA-derived templates, which is very efficient in processing large volumes of DESI spectra within a short time frame. However, this method occasionally yields unphysical model fits for galaxies and fails to adequately absorb calibration errors that may still be occasionally visible in the reduced spectra. Our proposed approach improves upon this existing method by refitting the spectra with carefully generated physical galaxy archetypes combined with additional terms designed to absorb data reduction defects and provide more physical models to the DESI spectra. We test our method on an extensive dataset derived from the survey validation (SV) and Year 1 (Y1) data of DESI. Our findings indicate that the new method delivers marginally better redshift success for SV tiles while reducing catastrophic redshift failure by $10-30\%$. At the same time, results from millions of targets from the main survey show that our model has relatively higher redshift success and purity rates ($0.5-0.8\%$ higher) for galaxy targets while having similar success for QSOs. These improvements also demonstrate that the main DESI redshift pipeline is generally robust. Additionally, it reduces the false positive redshift estimation by $5-40\%$ for sky fibers. We also discuss the generic nature of our method and how it can be extended to other large spectroscopic surveys, along with possible future improvements.

\end{abstract}

\keywords{Galaxy spectroscopy(2171), Astronomical methods(1043), Redshift surveys(1378), Astronomy software(1855), Astronomy data analysis(1858)}


\section{Introduction} \label{sec:intro}

The Dark Energy Survey Instrument (DESI) is a Stage-IV large spectroscopic survey \citep{levi2013,DESI2016-Overview} that will observe more than $\sim 40$ million spectra of galaxies, stars, and quasars in the wavelength range $3600-9800$ \angstrom with an average spectral resolution ($R=\lambda/\Delta \lambda$) of $2000$ at the shortest wavelengths to $5500$ at the longest wavelengths \citep{DESI2022-Instrument,silber2023,miller2023}. The main survey started in May 2021 and will continue for five years, resulting in the largest 3D map of the Universe ever. The early spectroscopic data of more than $\sim 1$ million galaxies, quasars, and stars was released in early June 2023 \citep{desiedr2023}. The DESI early data release\footnote{EDR is publicly available at \url{https://data.desi.lbl.gov/doc/releases/edr/}} (EDR) consists of raw images, reduced spectra, spectral classifications, and redshifts of each target observed with DESI during the survey validation (SV) phase. This requires \tb{a large number of} software development to process such large datasets efficiently and has motivated the development of classical machine learning and neural network-based approaches to analyze the data. 

\modify{One of the principal aims of these surveys is to construct an accurate 3D map of the large-scale structure of the universe with these tracer galaxies and quasars. These maps allow us to measure baryon acoustic oscillations (BAOs) and the redshift-space distortions (RSDs) that help us understand the expansion history and growth of structure(s) in the Universe. This requires precise redshift measurements for galaxies and quasars as understanding the systematics in redshift efficiency is crucial for accurately modeling the number density of tracers that contribute to these signals. These systematics also impact the total error budget of the best-fit cosmological parameters \citep[see][for DESI systematics]{krolewski2024,yu2024}.}

\modify{At the same time, precise redshift measurements are also important for non-cosmological scientific analyses. For example, stacking observed spectra to detect the radio H\textsc{I} line from radio surveys \citep{anand2019} or other weak emission lines \citep{maddox2013} that are typically undetectable in individual spectra relies on accurate redshift information. Similarly, robust redshift estimates help reduce the misidentification of galaxy groups, which usually have similar redshifts differed only by their peculiar velocities \citep{wang2020}. Furthermore, very precise redshifts are necessary to model and understand the gas kinematics in the halos of galaxies using quasar absorption lines \citep{tumlinson2017}. Owing to the importance of precise redshifts in astronomy, it is imperative to develop fast, optimal, and robust redshift measurement pipelines for the large ongoing and upcoming spectroscopic surveys.}

Given its straightforward and simple mathematics, principal component analysis (PCA) has demonstrated remarkable success in astronomy. The method's popularity stems from its deterministic and quickly convergent nature when solving for linear coefficients. Recent progress has allowed us to improve PCA-based spectral fitting algorithms, incorporating astronomical data uncertainties while constructing PCA templates \citep{tsalmantza2012, bailey2012}. These templates have been used to model astronomical spectra of galaxies, stars, and quasars and measure their redshifts. An example of such a method is the SDSS redshift pipeline \citep{bolton12}, employed for modeling SDSS BOSS/eBOSS spectra and measuring their redshifts. It has worked extremely well on millions of spectra, which helped the SDSS collaboration construct one of the largest 3D sky maps at the time \citep{ahumada2016}. At the same time, PCA templates have been combined with stellar population synthesis models to measure galactic properties from their spectra \citep[see][for details]{chen12}.

However, one issue with PCA-based templates is their inability to present the physical properties of galaxy spectra in their eigenvectors alone. One commonly encountered problem in PCA fits involves fitting noise as negative and unphysical fluxes, \tb{particularly in low signal-to-noise (S/N) spectra} that yield inaccuracies in measuring their redshifts and associated physical properties. This drawback is important to tackle in light of the ongoing large spectroscopic surveys that are poised to accumulate millions of spectra in the coming years.

To tackle these issues, several other modeling techniques have been used in recent years. One such method is Nonnegative Matrix Factorization (NMF), which is also a dimension reduction technique akin to PCA. It factorizes a large input matrix (typically the fluxes as a function of wavelength) into two smaller non-negative matrices\footnote{$V\approx WH$, where $W\geq 0$ and $H\geq0$ are found by minimizing the Frobenius norm $||V-WH||_{\rm F}$.}, namely eigenspectra, and coefficients, which can reproduce the input matrix fairly well. Furthermore, these eigenspectra exhibit a greater resemblance to the physical characteristics, like emission or absorption features in astronomical spectra. \citet{lee1999} presented a simple iterative update rule to derive these smaller matrices while ensuring a non-increasing nature for the associated cost function. While NMF presents inherent complexity as there is no optimal algorithm for finding the global minimum for the cost function associated with these two matrices, it is still extremely useful for astronomical spectra where the fluxes are always positive. Consequently, NMF has emerged as an alternative tool for modeling galaxy and quasar continua \citep[see][]{zhu13a, anand2021, napolitano2023}, especially to search for metal absorbers in quasar spectra in the context of circumgalactic/intergalactic media (CGM or IGM) or intracluster medium (ICM) studies \citep{anand2021, anand2022}. 

However, very few NMF algorithms exist that may work well with low S/N data. Recently, \citet{green2023} have extended the NMF algorithm to construct eigenspectra and coefficients for astronomical data with negative fluxes. This is an important result for low S/N spectra coming from low-resolution large spectroscopic surveys, where even carefully calibrated sky subtraction may result in negative fluxes in reduced spectra. 

\modify{An alternative method that has been previously employed involves utilizing correlation peaks in the cross-correlation function of the high-pass-filtered input spectrum with a series of templates. This technique was initially developed for the Galaxy And Mass Assembly (GAMA) survey galaxies, as described in \citet{baldry2014}. However, it is readily adaptable to forthcoming surveys such as DEVILS \citep{davies2021} on Anglo-Australian Telescope (AAT) and WAVES\footnote{Wide Area Vista Extragalactic Survey.} on the Vista telescope with 4MOST \citep{driver2016}. This approach offers two advantages over traditional PCA-based redshift determination methods. Firstly, it provides highly reliable redshifts with uncertainties as low as $<50$ \kms. The DESI's PCA-based redshift fitter also achieves an uncertainty of less than $40$ \kms in redshift measurements, similar to the correlation peak approach. Secondly, it facilitates the identification of multiple peaks in the cross-correlation function, which helps identify overlapping galaxies (such as those in clusters or groups, e.g., \citealt{holwerda2015, holwerda2022}) or strongly lensed systems (e.g., the SLACS program initiated by \citealt{bolton2008}).}

The recent progress in neural networks and machine learning methods have also played a crucial role in analyzing astronomical datasets, encompassing tasks from image and spectral classification to the prediction of cosmological parameters \citep[see][for a review]{baron19}. An example of such an approach is QuasarNET, a convolutional neural network (CNN)-based method developed for the classification and redshift measurement of quasars \citep{busca18} observed with surveys such as eBOSS and DESI. Consequently, the astronomical community relies on these approaches to analyze large numbers of spectra.

To address the problem of \tb{unphysical line fitting in PCA approach}, some studies have explored redshift fitting (using a minimum $\chi^{2}$ approach) of galaxy spectra using physical galaxy models (also known as the \textit{archetypes}) and polynomials \citep{cool13}. An example of such automated software is \package{redmonster} \citep{hutchinson2016}, which was developed for the eBOSS program of the SDSS. The software yielded a high redshift success rate ($\sim 90.5$\%) for luminous red galaxies (LRGs). However, this model had the freedom to take negative coefficients in order to achieve the optimal solution, which made it susceptible to unphysical modeling of negative features as real absorption features. Though it worked fine for early SDSS datasets, it was not computationally efficient and, therefore, not suited for a survey as large as DESI, which aims to observe 10 times more spectra than SDSS. 

In this paper, our objective is to improve the existing method for classifying and measuring redshifts of the DESI spectra. We aim to integrate the existing DESI redshift fitter with a suite of physically \tb{motivated} archetypes, combined with a set of polynomials to construct spectral fits that are more physical. This also allows us to maintain the efficiency intrinsic to PCA. We introduce several new features in our model, such as a new state-of-art set of archetypes, \tb{Legendre polynomials, Gaussian priors on polynomial coefficients}, and enforcing positive coefficients to the archetypes to avoid any unphysical fittings of the spectra. 

The structure of our paper is as follows: We describe the spectral data and DESI-specific methods and techniques in section~\ref{desidata}. In Section~\ref{method}, we state the problem and illustrate our methodology and implementation. Section~\ref{code_performance} presents extensive tests and the performance evaluation of our method using existing DESI data. Finally, in Section~\ref{discussion}, we discuss the implications of our method and potential future avenues of exploration. 

\begin{figure*}
    \centering
    \includegraphics[width=0.8\linewidth]{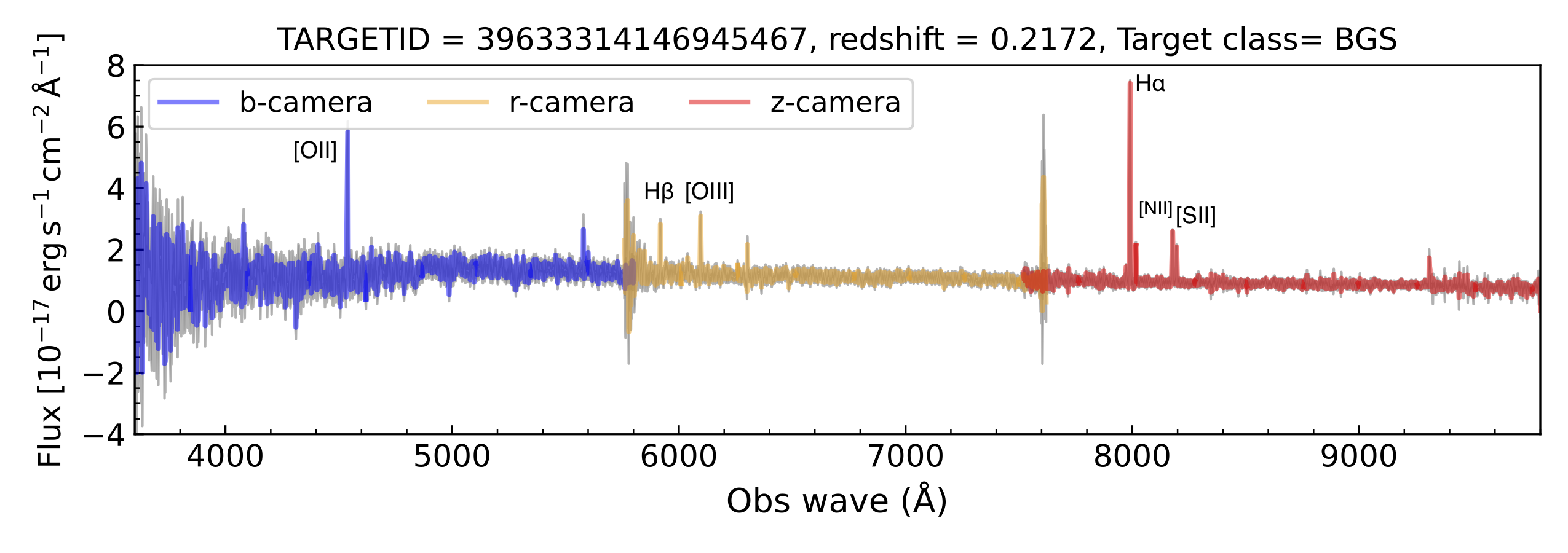}
    \includegraphics[width=0.8\linewidth]{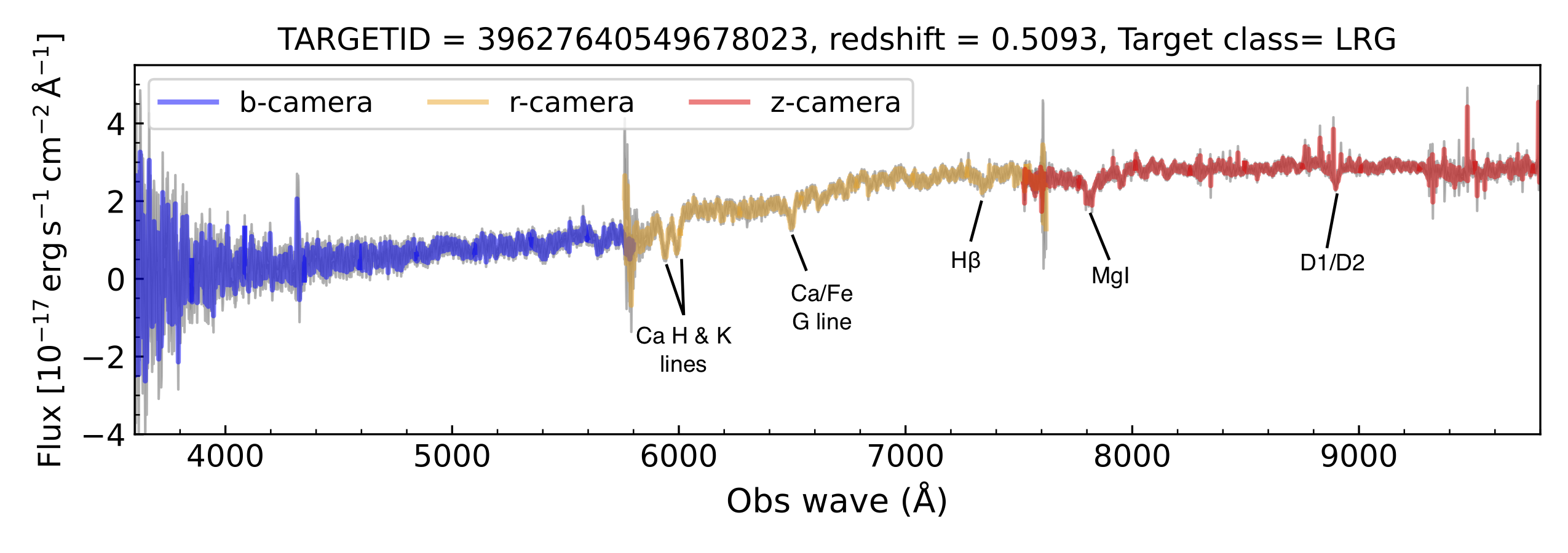}
    \includegraphics[width=0.8\linewidth]{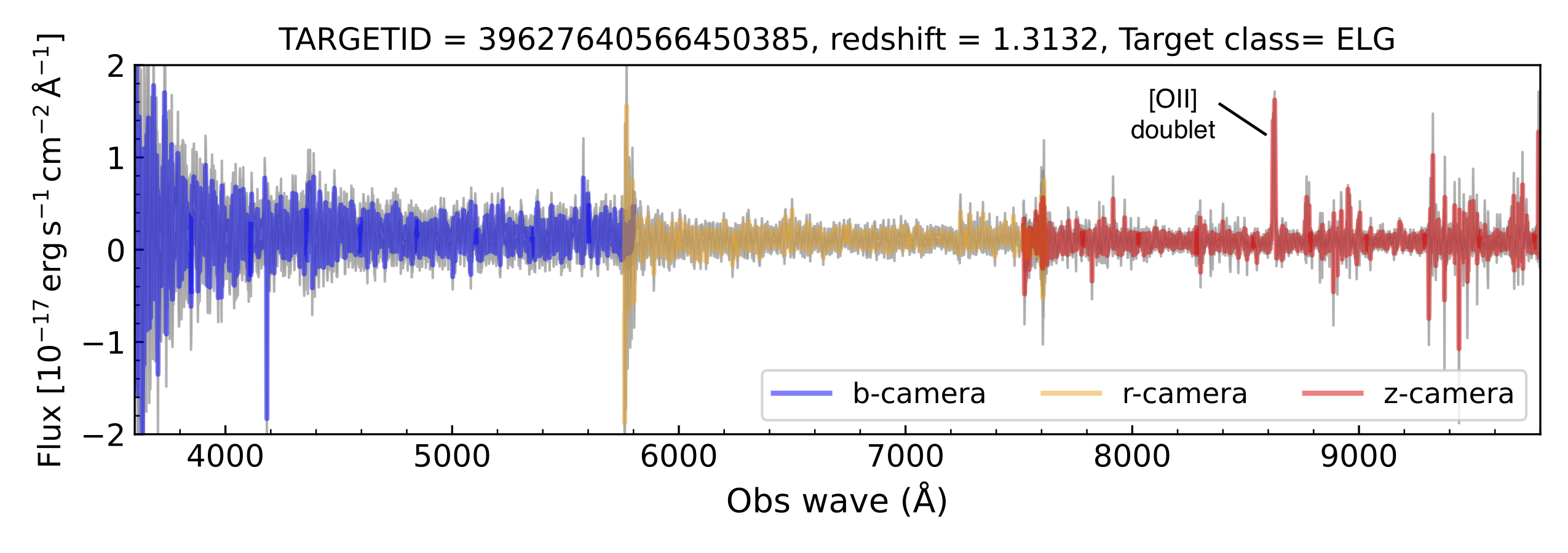}
    \includegraphics[width=0.8\linewidth]
    {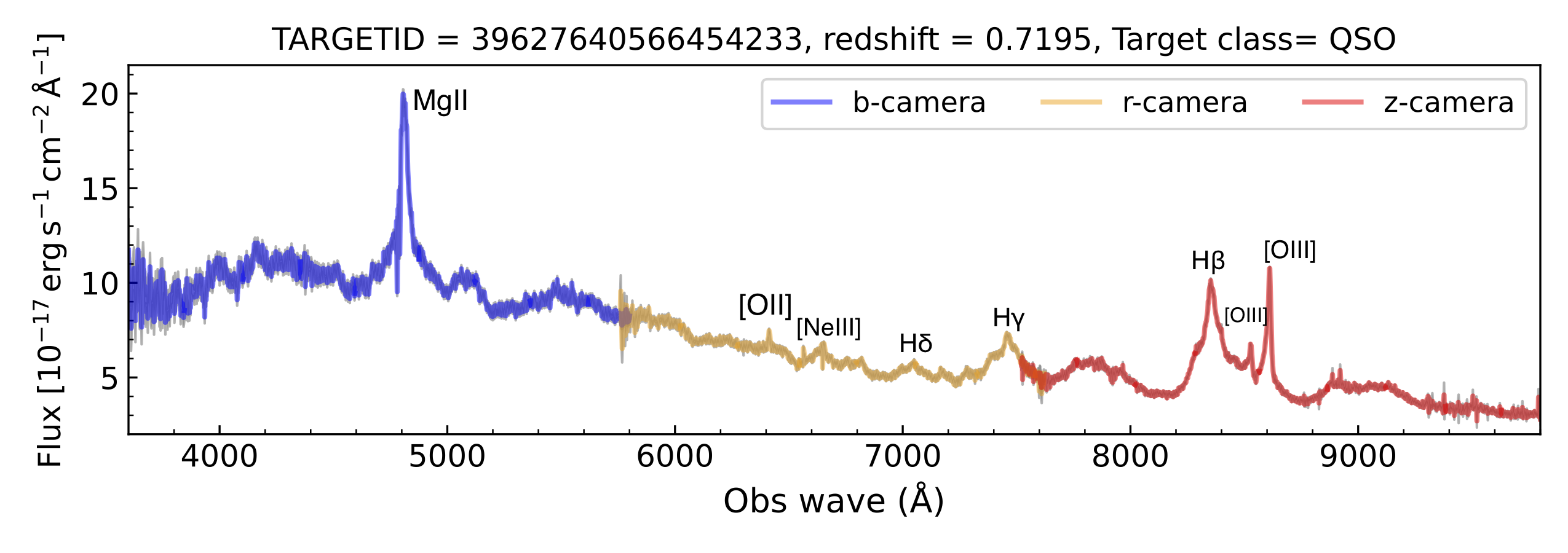}
    \caption{DESI example spectra of BGS, LRG, ELG, and QSO targets. The colored lines show the three DESI cameras. The uncertainties in fluxes are shown in gray in each panel. We also label the expected location of absorption and emission lines in each spectrum.}
    \label{fig:desispectra}
\end{figure*}

\section{DESI Data And Classification Procedure}\label{desidata}

\subsection{DESI Spectra}\label{desispectra}

The DESI focal plane is segmented into $10$ wedges or petals, each connected to a dedicated spectrograph. These spectrographs consist of three cameras denoted as $b$ (3600--5800~\AA{}), $r$ (5760--7620~\AA{}), and $z$ (7520--9824~\AA{}), each exhibiting a resolving power ranging from $2000$ to $5000$. Each petal is equipped with $500$ fibers, each directed towards unique sky positions. Consequently, DESI can simultaneously observe spectra from $5000$ targets. \tb{The fiber assignment\footnote{\tb{The fiber assignment algorithm basically selects which fibers are to be assigned to the targets, blank sky location, and calibration stars for each tile. The priority of targets means they will be assigned first or last while assigning the fibers. For e.g., QSO targets will always be assigned first on dark tiles followed by LRG and ELG targets \citep{DESI2022-Instrument}.}} algorithm is run throughout the night on the fly, and specific targets (along with their RA and DEC) are assigned to each of the 5000 fibers, constituting a single observing "tile."} Furthermore, a subset of the 5000 fibers (usually $\sim 20$) is specifically allocated for observing blank sky regions to measure the sky emission model for spectral reduction.  These designated fibers are referred to as "sky fibers." \modify{These are pre-selected to target specific regions of the sky that are empty. The sky fibers are the same for each tile, though they can point in different directions for each tile.} \tb{These tiles} are designed for specific observation programs, the two main ones being the \textit{dark} and \textit{bright}. As their name indicates, they are optimized for different observation conditions, and the observations switch from one to the other depending on a combination of moon brightness, $r-$band sky magnitude, sky transparency, and seeing \citep[see][for more details]{myers2022, schlafly23}. Moreover, certain targets are observed multiple times, and their spectra are combined across exposures to construct very high S/N spectra.

The instrumental design of DESI allows it to observe many galaxies and quasar targets with broad physical properties, such as stellar populations, emission lines, and dust content over a wide range of redshifts. DESI targets three broad classes of galaxies: bright galaxy sample (BGS,  \citealt{hahn2023BGS}), luminous red galaxies (LRGs, \citealt{zhou2023LRG}), and emission-line galaxies (ELG, \citealt{raichoor2023ELG}). \tb{Galaxies in the DESI BGS sample are low redshift, and they span from massive and quenched galaxies with evolved stellar populations to low-mass and star-forming galaxies with young stellar populations.} They are further classified as \target{BGS\_BRIGHT} and \target{BGS\_FAINT} depending on their brightness and magnitudes \citep{myers2022}. Galaxies in the DESI BGS sample are observed in bright conditions on \textit{bright} tiles. Next, the LRGs are targeted based on their $grzW1$ photometry and 4000~\AA\, break \citep[see][for detailed selection cuts]{zhou2023LRG} and are dominated by a metal-rich stellar population and low star-formation rates. They are also more massive than BGS and targeted in the redshift range of $0.4<z<1$. 

Next, the ELG sample includes primarily star-forming galaxies, and their spectra show several high equivalent width emission lines. The principal emission line that is aimed to resolve for ELGs in the DESI survey is the [\ion{O}{2}]~$\lambda\lambda3727,29$ doublet. They are targeted in the redshift range of $0.6<z<1.6$ \tb{with three main selections: \target{ELG\_LOP}, \target{ELG\_HIP} and \target{ELG\_VLO}.} The main difference between \target{ELG\_LOP} and \target{ELG\_VLO} is based on $g-r$ and $r-z$ cuts such that these two sets are disjoint and target different redshift bins. \target{ELG\_LOP} has higher priority in fiber assignment and covers galaxies in $1.1<z<1.6$, where other DESI tracers are less dense, while \target{ELG\_VLO} has lower priority and targets in $0.6<z<1.1$ \citep[see][for more details]{raichoor2023ELG}. \tb{\target{ELG\_HIP} is a 10\% random subsample of \target{ELG\_LOP} and \target{ELG\_VLO} and has the same fiber assignment priority as LRGs.} DESI has also been designed to observe millions of quasars in the redshift range $z>1.6$ to understand the large-scale structure and constrain the cosmological parameters. The QSOs are targeted with optical and optical+IR color cuts in $g-z$ vs $grz-W1$ color space \cite[see][for more details]{chaussidon2023}. \tb{QSOs have the highest fiber assignment priority followed by LRGs, \target{ELG\_LOP} and \target{ELG\_VLO}. These targets are observed in dark times on \textit{dark} tiles only.}

\begin{table}
\centering
\caption{DESI Target selection, where $grz$ are magnitudes from Legacy Survey (LS). References: [a] \citet{hahn2023BGS}, [b] \citet{raichoor2023ELG}, [c] \citet{zhou2023LRG}, [d] \citet{chaussidon2023}. $g_{\rm fiber}$ and $z_{\rm fiber}$ are the $g$ and $z-$ band fiber magnitudes, i.e., the magnitude corresponding to the expected flux within a DESI fiber.}
  \addtolength{\tabcolsep}{-1.25mm}
  \begin{tabular}{cccc}
    \hline
    Target class& Magnitudes &Redshift& Refs.\\
    \hline
    \target{BGS\_BRIGHT}&$r<19.5$&$<0.4$&a\\
    \target{BGS\_FAINT}&$19.5<r<20.175$&$<0.4$&a\\
    \target{ELG\_LOP}&$g_{\rm fiber}<24.1$&$\in (1.1,1.6)$&b\\
    \target{ELG\_VLO}&$g_{\rm fiber}<24.1$&$\in (0.6,1.1)$&b\\
    \target{LRG}&$z_{\rm fiber}<21.6$&$\in (0.4,1)$&c\\
    \target{QSO}&$r<23$&>1.6&d\\
    \hline
  \end{tabular}
  \label{tab:target_selection}
\end{table}

\modify{The apparent magnitudes and redshift range of the main DESI targets are compiled in Table~\ref{tab:target_selection}.} We also present the typical DESI spectra in their observed frame corresponding to each target class in Figure~\ref{fig:desispectra}. The top panel shows the spectrum of a representative low-redshift BGS galaxy featuring multiple emission lines. The second panel displays the spectrum of an LRG, showing the clear Ca \texttt{H} and \texttt{K} absorption features and 4000~\AA\ break (redshifted at $\sim 6000$~\AA). An example spectrum of ELG is presented in the third panel, highlighting the clearly visible \oii doublet line redshifted at $\sim8600$~\AA. Finally, in the last panel, we present a typical DESI quasar spectrum, where broad emission lines (hydrogen and metal lines) are distinctly visible on the top of the quasar power-law continuum.

DESI observations are divided into two categories: survey validation (SV) phase and main survey (MS). The SV phase ran between Dec 2020 and May 2021, during which several tiles were observed, details of which are fully described in \citet{myers2022}. The SV tiles were designed to observe distinct target classes (e.g., LRG, BGS, ELG, QSO) on several nights encompassing different exposure times (hence varying S/N), weather, and observation conditions. This comprehensive dataset provides a unique opportunity to thoroughly assess and quantify the performance of the DESI instruments and the spectral reduction and redshift estimation pipeline. The SV phase full datasets were publicly released as part of DESI EDR. Additional details regarding these observations can be found in \cite{desisv2023}. The main survey started in May 2021 and will continue for five years until May 2026. The dataset will be publicly released by the DESI collaboration. The Year 1 (Y1) data of the main survey that encompasses observations spanning approximately $\sim 1$ year, from May 2021 to June 2022, will be released first. 

In this manuscript, we use targets observed during the SV phase and Year 1 to assess the efficacy of our archetype-based method for redshift fitting specifically for galaxy spectra. To achieve this, we carefully constructed a set of "galaxy archetypes" (described in section~\ref{archetypes}) and applied our approach (described in section~\ref{method}) to DESI targets. We quantitatively evaluate the performance of our algorithm and provide and shed light on its effectiveness and potential impact, as detailed in the subsequent sections. 

\subsection{Current DESI Redshift Fitter}\label{redrock_detail}

Redrock\footnote{\url{https://github.com/desihub/redrock/}} \citep{bailey2024} is the principal spectral fitter and redshift estimator software for the DESI spectra. It is very computationally efficient and can be run on both CPUs and GPUs. Though Redrock is run specifically on DESI spectra, the methods and implementation are quite generic. The underlying algorithm selects the least $\chi^{2}$ from PCA-based templates fit over a range of redshifts for three main spectral types: galaxy, stars, and QSOs. The QSO and stellar templates are further divided into different subtypes to cover their diversity. The redshift and spectral class solution corresponding to the lowest $\chi^{2}$ is the final redshift and spectral type of the input spectrum. The templates were generated from a combination of real and synthetic spectra of astronomical targets using an iterative principal component generator, \textit{empca} \citep{bailey2012}, which also takes uncertainties of the data into account. 

\citet{bailey2024} provides the details of each template and its performance. The principal components for galaxies were generated using a set of 20,000 synthetic spectra (described in section~\ref{galaxy_spectra}). In contrast, the QSO templates were constructed using quasars observed with the eBOSS program of SDSS \citep{brodzeller2023}. On the other hand, for stars, the PCA templates were generated using synthetic stellar spectra in six different $T_{\rm eff}$ (corresponding to stellar subtypes: \textit{B, A, F, G, K} and \textit{M}) bins to account for the broad diversity among stars \citep{allendeprieto2018, cooper2023}. In addition, separate PCA templates were generated for cataclysmic variables (CVs) and white dwarfs (WDs) as their physical properties differ significantly from main-sequence stars. The CV and WD templates were generated using the same archetype technique described in \citet{bolton12}.

Redrock is very efficient and has performed well on SV and Y1 datasets. However, it occasionally yields negative flux models for some low S/N spectra, where it fits noise or poorly subtracted sky signals as negative features (see section~\ref{archetypefit}). Currently, if the method identifies negative flux at the location of a forbidden line, such as \oii, it implements an ad-hoc method to correct these unphysical fits. \tb{It adds a prior to the $\chi^{2}$ based on the model flux\footnote{\tb{$[\rm OII]_{\rm model\,flux}$$ = |\sum_{k}a_{k}T_{k}(\lambda_{i})|, \, 3724<\lambda_{i}<3733$, where $\lambda_{\rm i}$ is rest-frame wavelength (around \oii line) in \AA\, and $T_{k}$ and $a_{k}$ are PCA templates and coefficients.}} around the \oii line.} This correction forces the final $\chi^{2}$ to be higher and modifies the ranking of other models. However, this lacks a solid mathematical basis and does not work for other emission lines in the current version (\package{0.19.0})\footnote{\url{https://github.com/desihub/redrock/releases/tag/0.19.0}}.

Redrock also lacks algorithmic features capable of accurately absorbing the calibration errors that may occasionally show up in the reduced spectra. It is difficult to quantify the occurrence of such spectral artifacts (described below) as Redrock does not raise any warning bits for such issues. However, they still need attention, as even a relatively small failure rate can be large in absolute numbers, given the survey size. 

There are two \tb{primary calibration issues} \citep[see][for more detailed discussion]{guy2023} considered in this work. First, the slightly different bias or zero level in the CCD quadrants causes discontinuity at the boundaries, from where the electrons are read by the amplifiers present at all four corners (see Figure 2, 55 and Appendix E of \citealt{guy2023}). Similar discontinuities can arise from fluctuations of dark current for some CCD columns. This causes the entire spectrum to shift up or down in one or more DESI cameras (more precisely, at half of the camera), corresponding to one of the amplifiers for some of the fibers). These discontinuities can still show up even after the flux calibration as these effects are additive, and the flux calibration is multiplicative. The flux calibration is obtained with bright standard stars, so the relative effect of an offset is barely visible. 

\tb{Secondly, the flux calibration corrects for the difference in throughput from one camera to another. However, as described in sections 4.6.3 and 4.6.9 of \citet{guy2023}, there remain wavelength and focal plane position-dependent calibration errors caused by the chromatic distortions of the corrector (see also Figure 39, \citealt{guy2023}). Fibers are positioned to get the best throughput in the $r-$band at the cost of higher losses at other wavelengths. For this reason, we consider additional corrections beyond a pure offset for redshift fitting. In general, this correction is less significant than the first one.} 

To address these issues in the current DESI redshift fitting pipeline, we propose a different approach. Our proposed method combines PCA-based redshift scanning with physically motivated galaxy spectra (\textit{archetypes}, more details in section~\ref{archetypes}) to model the DESI spectra and measure the redshifts. Additionally, we introduce polynomial modeling in each camera to accurately model these calibration errors and absorb them in the best-fit redshift model. We provide the mathematical description of our model in the next section, followed by an extensive demonstration of how it naturally resolves these issues and performs on the DESI survey validation phase and Y1 datasets. 

\section{Method} \label{method}

\begin{figure*}
    \centering
    \includegraphics[width=0.95\linewidth]{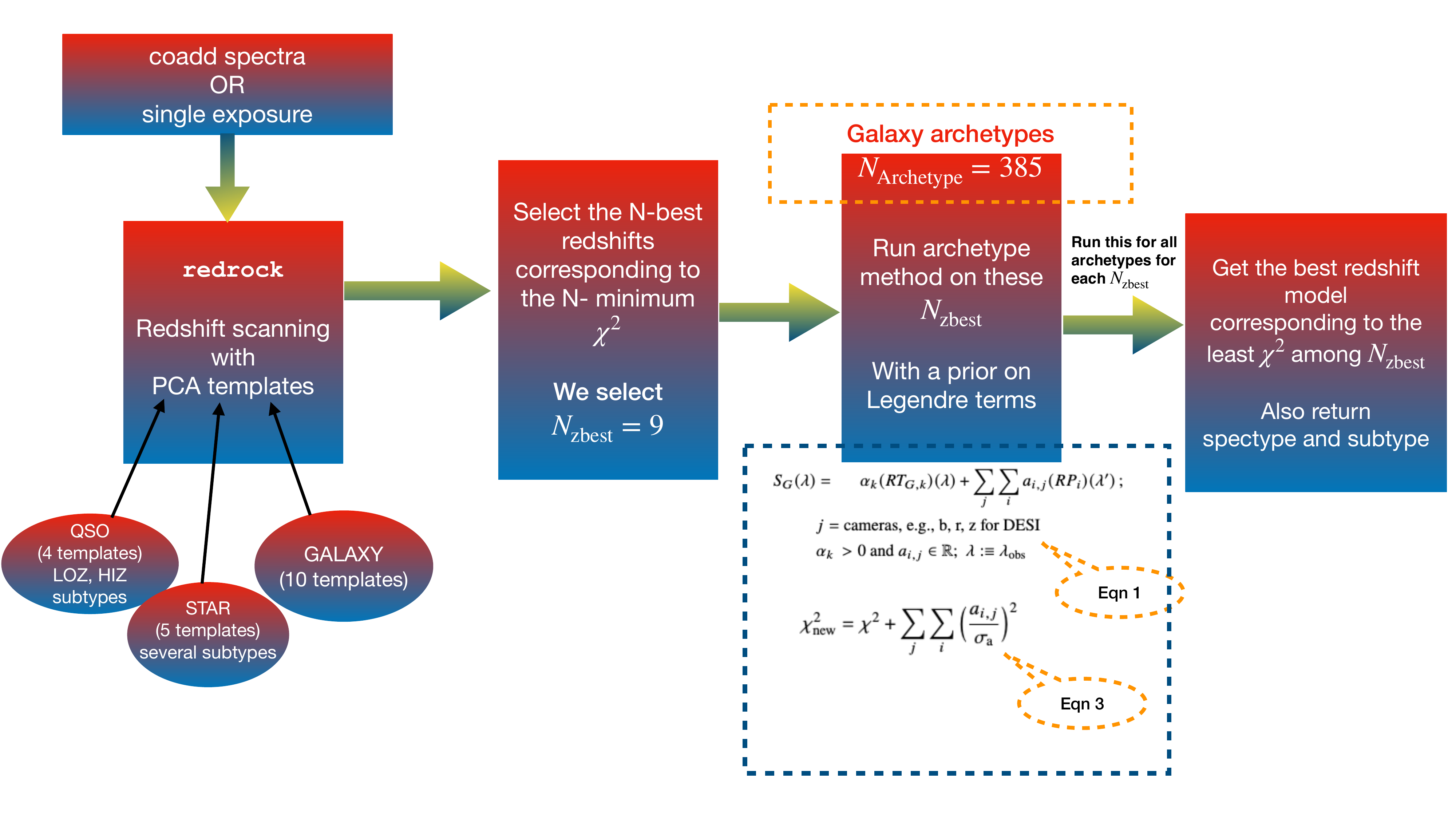}
    \caption{A schematic of our archetype based per camera polynomial fitting. The coadded or single exposure spectra are input to the redrock that uses previously estimated PCA-based templates of several spectral types to perform the redshift scan. We retain the $N_{\rm zbest}$ redshifts corresponding to the least $N$ $\chi^{2}$ values after the preliminary scan. Subsequently, we model the input spectra using Eqn~\ref{eqn:per_camera_model} at each of these $N_{\rm zbest}$ redshift (in a loop), employing each archetype (in another loop) to determine the final best $z$ associated with the least $\chi^{2}$. The final output is the best-fit redshift, along with the spectral type and index of the best-fit archetype.}
    \label{fig:method_flowchart}
\end{figure*}

\subsection{Modeling Galaxy Spectra}\label{archetype_model}

As elaborated in Section~\ref{redrock_detail}, Redrock suffers from certain challenges. There are three main issues we are trying to resolve with our new approach, as detailed in the previous section. 1) unphysical fitting of negative fluxes in DESI spectra, 2) discontinuities in the spectrum due to CCD bias and zero issues, and 3) \tb{gradient-like throughput offsets caused by chromatic distortions of the corrector}. In the current formulation, we use Legendre polynomials to model these two spectral defects \tb{and archetypes to yield more physical models for the spectrum. Given that our archetypes span the parent galaxy sample fairly well (more discussion in section~\ref{archetypes}), we model the input spectrum with a single archetype. This is mathematically less complex and also computationally easier to implement.} Finally, our model can be written as follows:
\begin{equation}
\begin{split}
    S_{G}(\lambda) &= \alpha_{k} (RT_{G, k})(\lambda) + \sum_{j}\sum_{i}a_{i, j}(RP_{i})(
    \lambda '
    ) \,;\,\,\,\, \\&j= \text{cameras, e.g., } b,\, r,\, z \text{ for DESI} \\& \alpha_{k}\,>0 \text{ and }a_{i,j}\in \mathbb{R};\,\, \lambda: \equiv \lambda_{\rm obs} \\ 
    \end{split}
    \label{eqn:per_camera_model}
\end{equation}
where $\lambda$ is observed wavelength, $T_{G, k}(\lambda)$ are $k^{\rm th}$ galaxy archetype, $P_{i}(x)$ are Legendre polynomials with degree $i$, $\lambda'$ is reduced wavelength\footnote{\tb{$\lambda' = 2\cdot \frac{\lambda - \lambda_{\rm min}}{(\lambda_{\rm max}-\lambda_{\rm min})} -1 \in [-1, 1]$, where $\lambda_{\rm min}$ and $\lambda_{\rm max}$ are minimum and maximum observed wavelengths in each band.}} and $\alpha_{k}$, $a_{i, j}$ are linear coefficients for archetypes and Legendre polynomials, respectively and $R$ is the resolution matrix\footnote{It is used to convert the correlated fluxes into uncorrelated ones: $F' = RF$, $F'$ is final flux that is used in calculating $\chi^{2}$.}, which is used to decorrelate the data and accounts for the resolution of the spectrographs \citep[see][for more details]{guy2023}. Applying the $R$ matrix on templates is a relatively slower step in Redrock as it involves a large number of matrix operations. Note that $\alpha_{k}>0$ is fit over entire spectra, so for each spectrum, \tb{there is just one archetype coefficient.} At the same time, there are $3(d+1)$ coefficients (\tb{$3$ for three cameras in DESI}) for polynomial terms where $d$ is the highest degree of polynomial used in modelling. This fitting approach is applied to all the cameras simultaneously, ensuring accurate modeling of spectra in each of the wavelength bands. In total, our model has $1+3(d+1)$ coefficients. 

Moreover, we also want to leverage the efficiency and computational speed of PCA and fit the spectra; therefore, after scanning for redshifts using PCA templates, we select the N- best redshifts (or $N_{\rm zbest}$, corresponding to the N-least $\chi^{2}$ (see Figure~\ref{fig:method_flowchart} for the method schematic) \tb{regardless of redrock spectral classifications} and refine the model fit for each redshift with a linear combination of physical spectra (to model the shape and features of spectra) and Legendre polynomials (to model pipeline errors), which can result in different ranking in terms of best fit $\chi^{2}$. \tb{Here, it is important to understand that our model is not independent of current redrock as we run our method on a list of redshifts ($N_{\rm best}$) obtained with the initial PCA-based method. Therefore, if the true redshift does not fall within this list, our method will also fail. Therefore, the choice of $N_{\rm zbest}$ is also a free parameter in our method, and we explored this space (see section~\ref{more_tests}) to achieve an optimal value. \modify{We also emphasize that the statistical precision of the redshifts in the archetype method will be the same as that of the Redrock. Our focus in the paper is more on improving catastrophic failures than redshift accuracy.} We aim to perform an independent test of the archetype procedure as a redshift classifier in the future.}

Next, to model the CCD discontinuities and throughput offsets in the spectra, we use only the first two Legendre polynomials, i.e., a constant and slope term ($a_{j,0}, a_{j,1}$). However, the code can accommodate several Legendre terms, and we could explore this possibility in the future. These per-camera coefficients are also fit simultaneously. Finally, we have seven coefficients (a principal archetype and one pair of Legendre coefficients for each of the three cameras). We then solve for the coefficients with \texttt{scipy's} \texttt{optimize.lsq\_linear}\footnote{\url{https://docs.scipy.org/doc/scipy/reference/generated/scipy.optimize.lsq\_linear.html}} module. To summarize, this algorithm solves for the coefficients subject to the specified bounds (described in Eqn~\ref{eqn:per_camera_model}) through an iterative least-square solving method known as \textit{bounded-value least-square} (\bvls), initially proposed by \citet{stark1995}. \textit{BVLS} has been shown to converge to a solution in a similar way the nonnegative least squares (NNLS) converge, as it is modeled on the NNLS iterative approach as developed by \citet{lawson1976}. We use this method described above and choose the final redshift that corresponds to the least $\rm \chi^{2}$ (see Figure~\ref{fig:method_flowchart} for detailed schematic). For a spectrum given its flux, $F(\lambda_{\rm i})$ and variance $\sigma^{2}(\lambda_{i})$, the $\rm \chi^{2}$ from its best fit model ($S_{G}$) is calculated as 

\begin{table}
\centering
  \caption{Redshift success rate criteria ($Q_{\rm o}$) for spectra having nominal exposure time, i.e. for bright tiles, $T_{\rm eff}>180\rm s$ and for dark tiles, $T_{\rm eff}>1000\rm s$. $\rm OII_{SNR}$ is signal-to-noise ratio of \oii flux, which also requires that $\rm [OII]_{flux}>0$ and corresponding $\rm \sigma_{[OII]_{flux}}>0$, and $f(\Delta \chi^{2}) = 0.9-0.2\cdot  \rm log_{10}(\Delta\chi^{2})$. \target{ZWARN=0} is the bit warning for a redshift without any obvious issue. References: [a] \citet{hahn2023BGS}, [b] \citet{zhou2023LRG}, [c] \citet{raichoor2023ELG}, [d] \citet{chaussidon2023}.}
  \addtolength{\tabcolsep}{-1.25mm}
  \begin{tabular}{ccc}
    \hline
    Target& Redshift Selection criteria & Refs.\\
    &Common condition: \package{COADD\_FIBERSTATUS=0}\\
    \hline
    BGS&\package{ZWARN=0}, $\Delta \chi^{2}>40$&a\\
    LRG&\package{ZWARN=0}, $\Delta \chi^{2}>15$, $z<1.5$&b\\
    ELG&\package{ZWARN=0}, $\rm log_{10}(OII_{SNR})$$>f(\rm \Delta\chi^{2})$&c\\
    QSO\footnote[9]{\tb{SPECTYPE is from Redrock. This is a very preliminary selection criterion to define the redshift success rate for QSOs to compare with our archetype results. Our goal here is to just understand if the archetype recovers the QSO targets as QSOs.} For cosmology purposes, the good redshift for QSOs is defined based on a combination of redshifts obtained with afterburners that use redrock outputs as priors to estimate refined redshifts using a CNN and emission line modeling. More details on this can be found in \citet{chaussidon2023}.}&\package{SPECTYPE=QSO} &d\\
    \hline
  \end{tabular}
  \label{tab:z_success}
\end{table} 

\begin{equation}
\centering
\begin{split}
\chi^{2} &= \sum_{i}\Big(\frac{F(\lambda_{\rm i})-S_{G}(\lambda_{i})}{\sigma(\lambda_{\rm i})}\Big )^{2}; \,\,i = \text{ wavelength pixels}
\end{split}
\label{eqn:chi2}
\end{equation}

\tb{Finally, If a spectrum is typical of galaxies and there are pipeline defects in any of the cameras, the new model should have a lower $\chi^{2}$ than the best PCA fit (i.e. without archetypes and Legendre polynomial), as we allow Legendre polynomials to absorb these discontinuities. Hence, the final redshift should correspond to the archetype model rather than the PCA-based fit.} In contrast, if the spectrum is vastly different from a typical galaxy spectrum, our method should yield a larger $\chi^{2}$ than the PCA-only (without archetypes) model. 

The model $\chi^{2}$ is calculated for each archetype at each redshift (from best $N_{\rm zbest}$ redshifts) in a loop, so there are $N_{\rm archetype} \times N_{\rm zbest}$ loops in total. Then, we select the final redshift of the input spectra corresponding to the least $\chi^{2}$ \tb{and also store the corresponding archetype subtype (ELG, BGS, or LRG)} as the best-fit model for the input spectrum, in case the archetype $\chi^{2}$ is smaller than PCA-only (without archetypes) model. The final output redshift file also includes a $\Delta \chi^{2}$, the difference between the best-fit $\chi^{2}$ and second best-fit $\chi^{2}$. This difference in $\chi^{2}$ is a key indicator of a good redshift as it tells if the second best-fit model is significantly far from the best solution.

Using this $\Delta \chi^{2}$ parameter,  \textit{good redshift criteria} ($Q_{\rm o}$) is derived from an extensive analysis of DESI targets observed during the SV phase \citep[see][for more details]{desisv2023}. The conditions to define a reliable redshift are detailed in Table~\ref{tab:z_success}. Note that \target{COADD\_FIBERSTATUS==0} represents the spectra that are free from any hardware issue while \target{ZWARN==0} condition implies that redrock fits are good \citep[see][for more details]{desiedr2023}. We use the same criteria to assess the performance of our archetype-based redshift fitter. 

\begin{figure*}
    \centering
    \includegraphics[width=0.95\linewidth]{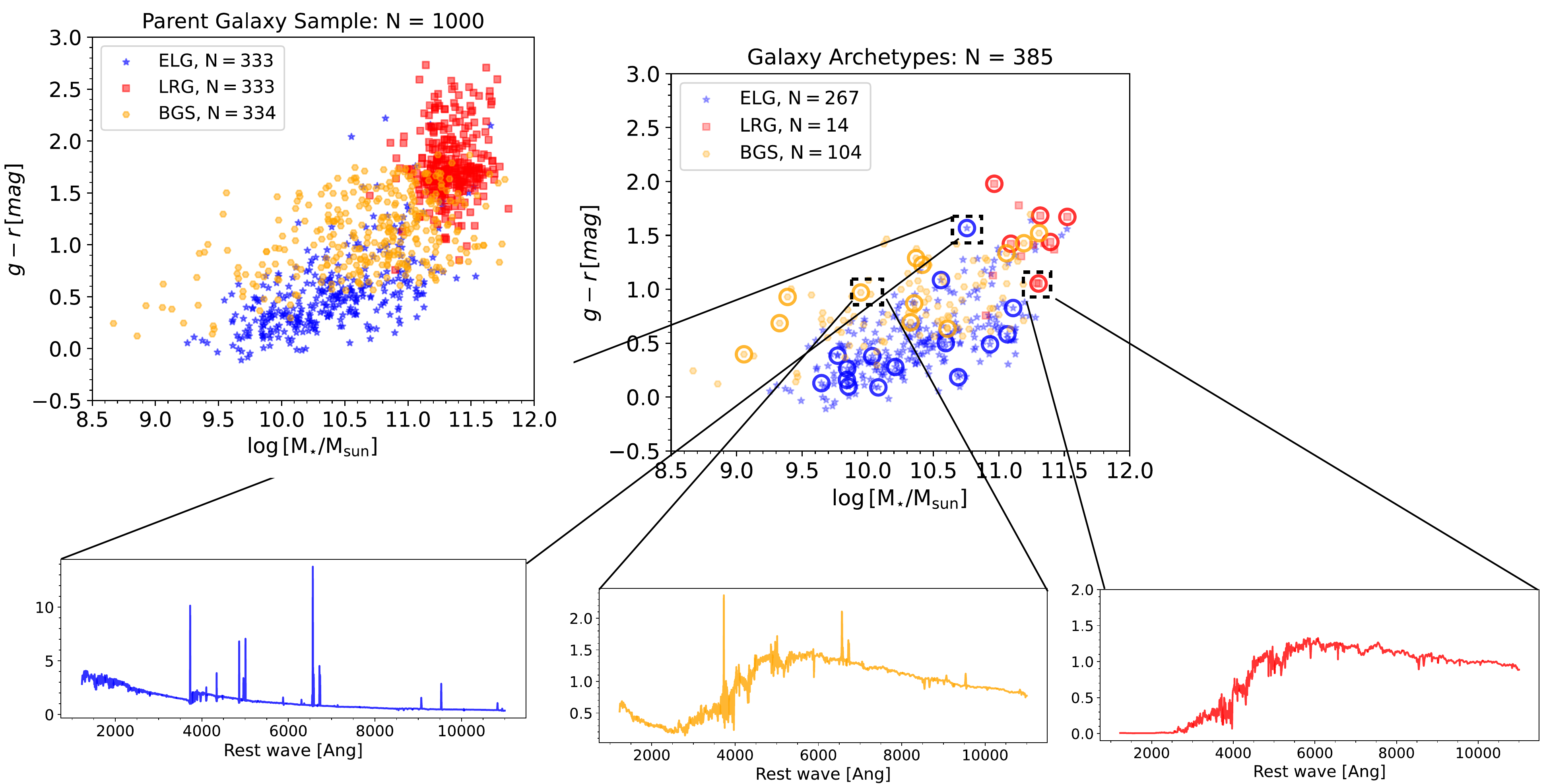}
    \caption{$g-r$ vs. stellar mass for parent galaxy sample (left) and archetypes (right). In both panels, we show the properties of ELGs (blue), LRGs (red), and BGS (orange). We see that galaxies lie in different regions on this plane, and archetypes also span a similar range in the color space. In the right panel, the open circles denote the archetype that we selected randomly to show their spectra in Figure~\ref{fig:example_archetypes} in  Appendix~\ref{archetype_vs_parent}.}
    \label{fig:archetypes}
\end{figure*}

\subsection{Priors on polynomial coefficients}\label{priors}

This section discusses the rationale for using priors on polynomial coefficients in the context of our current archetype-based redshift fitting method. The reason that we want to use priors is to avoid the correction for "multiplicative" broadband variations due to an incorrect broadband color of the model spectra, for instance, for QSOs. The challenge here arises from the fact that quasar spectra are often described by a power-law continuum, i.e., $F_{\rm c}\propto \lambda^{-\beta}$, with $\beta>0$, leading to a spectral slope. Unfortunately, this spectral slope can sometimes be incorrectly attributed to calibration issues within our proposed model. As a result, quasars can be misclassified as galaxies due to extreme freedom to our Legendre coefficients, which impacts spectral classification. We will highlight this more in section~\ref{zsuccess} when we calculate redshift success rates for all target classes. The use of priors aids in reducing the misclassification of genuine quasars as galaxies.

We want that polynomial correction should only correct for the "additive" terms caused by the CCD bias issue which we can understand better with sky fibers. By providing a user-defined prior ($\rm \sigma_{a}$), the algorithm adds a prior to the $\chi^{2}$ of Eqn~\ref{eqn:chi2_prior}, but only to the polynomial terms, before solving for their coefficients \tb{(see Figure~\ref{fig:method_flowchart} for the method schematic)}. \tb{This is similar to the Gaussian regularization that quantifies the prior by which the coefficients should be close to zero.} \tb{With the prior on polynomial terms, the }modified $\chi^{2}_{\rm new}$ takes the following form:
\begin{equation}
    \chi^{2}_{\rm new} = \chi^{2} + \sum_{j}\sum_{i}\Big (\frac{a_{i, j}}{\sigma_{\rm a}}\Big)^{2}
    \label{eqn:chi2_prior}
\end{equation}
\tb{It is important to note that the unit of prior is the same as the coefficients, which is proportional to the calibrated flux.}. 

Another key observation is that the default PCA galaxy templates result in artificially high $\Delta \chi^{2}$ values when fitting to sky spectra. While the priors avoid correcting quasar continuum variations (which are physical), they also help reduce plausible redshift estimates for sky fibers. These fibers, typically targeted to observe blank regions of the sky, should ideally yield negligible signal and thus low $\Delta \chi^{2}$. A small value of $\sigma_{a}$ adds a larger correction to the final $\chi^{2}$. 

For this study, we set $\sigma_{\rm a} = 0.1$, a value determined empirically based on our analysis of a subset of $339,712$ sky spectra from the Y1 dataset of DESI. We discuss this choice in more detail in Appendix~\ref{optimal_prior}, where we present the distribution of Legendre coefficients for skyfibers for both \textit{dark} and \textit{bright} tiles. Notably, the spread of the coefficient distribution varies between \tb{$0.1$ to $0.5$.} Our choice of $\sigma_{\rm a}=0.1$ closely aligns with this spread, ensuring that the application of priors also constrains the false positive redshift estimates for sky fibers. In summary, the priors on the Legendre coefficient are a valuable tool for reducing the misclassification of quasars as galaxies and for not yielding reliable redshift estimates for sky fibers. We show the results for both with and without priors in all the Tables and Figures in section~\ref{code_performance} and quantify the performance of our new method compared with Redrock (without archetypes). Finally, we show a detailed flowchart of all the steps of our method in Figure~\ref{fig:method_flowchart}.

\subsection{Synthetic Galaxy Spectra}\label{galaxy_spectra}

Prior to constructing a set of galaxy archetypes, it is important to carefully generate a set of synthetic galaxy spectra that closely align with DESI spectra in terms of wavelength coverage, resolution, and physical properties. The method to generate \textit {synthetic spectra} is based on a combination of observation and simulation, which is described in detail in \citet{bailey2024}. However, we provide a short summary of the approach below. We compile an extensive sample of galaxies at redshifts $0<z<2$ observed in different photometric and spectroscopic surveys. Then, we use physically motivated emission line fluxes and stellar population synthesis models to \tb{generate synthetic spectra} free of any instrumental effects of those surveys. Using these models, we generate very high-resolution spectra of galaxies in a rest-frame wavelength range so that the important emission and absorption lines fall within the DESI wavelength range. We describe the basic properties of these synthetic spectra below.

To generate \tb{synthetic spectra} for ELGs, we select galaxies (at $z\sim 1$) from the Deep Extragalactic Evolutionary Probe 2 (DEEP2) Galaxy Redshift Survey \citep{newman2013} Data Release 4 (DR4, \citet{ matthews2013}). The DEEP2 DR4 consists of $\sim 50,000$ high-resolution spectra of emission line galaxies that have measured [\ion{O}{2}]~$\lambda\lambda3727,29$ doublet lines \citep{newman2013} at an observed wavelength of $6500-9300$ ~\AA\, (typical \oii doublet region in DESI spectra for ELGs). ELGs are typically star-forming galaxies and appear blue in the image. The interaction of stellar activity (particularly the UV radiation from hot young stars) with the neighboring gas clouds can photoionize metals, which in turn can produce narrow forbidden metal lines that are visible on the top of the intrinsic stellar continuum in the galaxy spectra \citep[see][for review]{conroy2013, kewley2019}. The emission lines were generated based on theoretical models as described in \citet{stasinska2003} that can reproduce the observed trends in synthetic spectra. One caveat is that our emission line galaxy templates do not include any active galactic nuclei (AGNs), as they can show a combination of narrow and broad emission lines produced by complex physical processes due to central black holes and ongoing star formation. We will look into such subtypes in the future.

The BGS galaxy synthetic spectra were produced by using a flux-limited spectrophotometric sample of $11,000$ galaxies ($0.1<z<0.8$) observed in AGN and Galaxy Evolution Survey (AGES; \citet{Kochanek2012}. The spectral resolution and wavelength coverage are similar to DESI spectrographs.

Finally, we use a parent sample of $111,114$ LRGs observed with DECaLS/DR7 \textit{grzW1W2} photometry for the LRGs. At the same time, the redshifts were compiled from previous large spectroscopic surveys such as BOSS, AGES, and DEEP2 \citep{zhou2023LRG}. We aimed to simulate a physically motivated representative set of LRGs for targeting purposes and other DESI-related galaxy science. With this selected sample of LRGs, we generated synthetic spectra LRGs using \texttt{iSEDfit} package as described in \citep{moustakas2013a, moustakas2017a}. The method can fairly reproduce the observed SEDs of the LRGs. Note that our LRG modeling does not include emission lines\footnote{LRGs are usually passive and have no or very low star formation activity; therefore, they lack many emission lines.}, which may be important for a small fraction of DESI LRGs. A small fraction of the LRG targets can be AGNs that exhibit narrow metal lines in their spectra. 

Finally, with the public DESI code\footnote{\url{https://github.com/desihub/desisim}}, we can generate synthetic galaxy spectra whenever necessary. Using this package, we generated a sample of $1000$ (approximately 333 each) rest-frame ($\lambda=1228-11000$~\AA, $\Delta \lambda = 0.1$~\AA) synthetic spectra of LRGs, BGS, and ELGs. We present certain physical properties, specifically $g-r$ (color) vs. stellar mass, in the top left panel of Figure~\ref{fig:archetypes}. The colors blue, orange, and red correspond to ELGs, BGS, and LRGs, respectively. The galaxy population dichotomy \citep{kauffmann2003} is clearly visible where ELGs predominantly exhibit star-forming characteristics with a typical stellar mass of $\sim 10^{10.5}\, \rm M_{\odot}$, while LRGs are primarily passive and exhibit higher stellar masses ($\sim 10^{11.5}\,\rm M_{\odot}$). At the same time, the BGS sample occupies an intermediate position between ELGs and LRGs. 

\subsection{Galaxy Archetypes}\label{archetypes}

After generating these synthetic spectra, we use the classification technique, \textit{SetCoverPy}\footnote{\url{https://github.com/guangtunbenzhu/SetCoverPy}} \citep{zhu2016} to identify a subsample of spectra (aka \textit{archetypes}) that can sufficiently span the physical parameters of the parent sample. In a nutshell, it solves the Set Cover Problem (SCP), i.e., finding an optimal subsample that can represent the given parent sample using similarity (or distance metric) between the input spectra. A detailed description of both \textit{SetCoverPy} and SCP is beyond the scope of this paper; however, we encourage readers to refer to \citet{zhu2016} for a detailed discussion. It uses the \textit{Lagrangian relaxation} algorithm \citep{held1971, geoffrion1974, caprara99, Fisher2004-zl} to find the \tb{minimum} number of \textit{archetypes} that span the parent sample. \tb{One example of such use of generating archetypes using \textit{SetCoverPy} is \citet{brodzeller2022}, where it was used to identify quasar archetypes for purposes of creating physical models of quasar spectra.}

This method has two free parameters: the $s^{2}$ threshold (below which two instances or objects will be considered similar) on distance metric and the choice of weights as a function of wavelength. Once these two parameters are defined, \textit{SetCoverPy} finds the minimum number of \textit{archetypes}. We follow the same approach as described in section 4.3.2 of \citet{bolton12} to measure the similarity matrix of input spectra. We first calculate the \tb{distance metric}, $s_{ij}^{2}$ which measures the similarity between spectra $f_{i}$ and its model $am_{j}$, where $a$ is some scaling factor

\begin{equation}
\centering
s_{ij}^2=\sum_{l=1}^{N_{\rm pix}} \left[f_i\left(\lambda_l\right)- am_j\left(\lambda_l\right)\right]^2\
\label{eqn:s_chi2}
\end{equation}

$f_{i}$ corresponds to the normalized flux of $i^{\rm th}$ spectra. We normalize the flux in such a way that $\sum_{l=1}^{N_{\rm pix}}f_{i}^{2}(\lambda_{l})=N_{\rm pix}$. Note that we have used a constant uniform weight for all wavelengths (i.e., weights=1); therefore, they do not appear in the equation. As described in \citet{bolton12}, we can perform a least-square minimization on $s_{ij}^{2}$ (differentiating w.r.t $a$), which gives $a_{\rm best}(ij) = N_{\rm pix}^{-1}\sum_{l=1}^{N_{\rm pix}}f_{i}(\lambda_{l})m_{j}(\lambda_{l})$. With this, we finally get the minimum $s_{ij_{\rm min}}^{2}$

\begin{equation}
\centering
s_{ij_{\rm min}}^{2}=N_{\rm pix} \left(1-a_{\rm best}^{2}(ij)\right);\,\, \, s_{\rm red}^{2} = s_{ij_{\rm min}}^{2}/N_{\rm pix}
\label{eqn:scp_chi2}
\end{equation}

The rest-frame wavelength range we use in our $s_{ij_{\rm min}}^{2}$ calculation is $1228$ \AA\, to $11000$ \AA\, with $\Delta \lambda=0.1$~\AA,\, that gives us a total of $N_{\rm pix} =97,720$ \tb{wavelength pixels}\footnote{This is another dimension to explore in future. We chose this resolution in the paper as we want archetypes to have the same resolution as Redrock's PCA templates. However, this increases the file size.}.

A key point to note is that there is no optimal way to decide the best threshold ($s_{ij_{\rm min}}^{2}$) one should choose; it has to be decided empirically \citep[see][for more discussion]{zhu2016}. However, it can be understood from Eqn~\ref{eqn:scp_chi2} that a large threshold will only generate a small number of archetypes, which may not be a fair representation of the parent sample. In contrast, a small threshold will include similar archetypes and overrepresent the parent sample. \tb{It is important to run \textit{SetCoverPy} on all galaxies simultaneously without employing distinct thresholds for each galaxy type. Galaxy spectra occupy a large dimensional space lacking sharp boundaries in properties. Consequently, employing different thresholds for different types may not generate a set of diverse archetypes that span the galaxy properties space well.}

\begin{figure}
    \centering
    \includegraphics[width=0.975\linewidth]{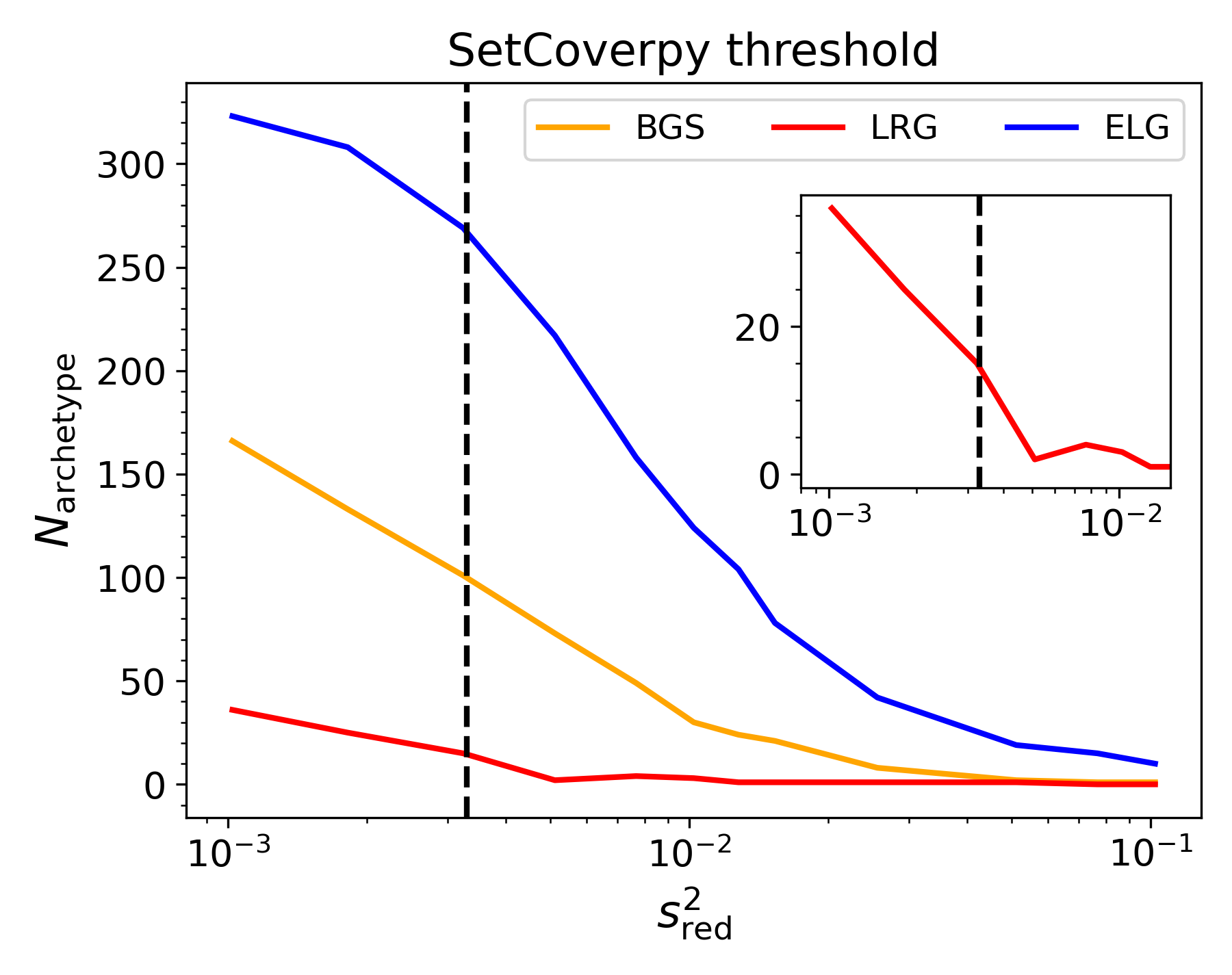}
    \caption{\modify{Empirical determination of optimal threshold for archetype generation using \textit{SetCoverPy} method. We show the $s_{\rm red}^{2}$ as a function of the number of galaxy archetypes for each galaxy subtype. Our optimal value for $s_{\rm red}^{2}\approx 0.003$ lies where the number of LRG archetypes significantly decreases (as shown in the inset). Beyond this threshold, the number of archetypes begins to saturate at higher $s_{\rm red}^{2}$ values.}}
    \label{fig:sred_square}
\end{figure}

Hence, we employ an empirical approach to determine our optimal selection. For our purpose, we used $s_{\rm red}^{2}$ value to decide the optimal threshold. We varied the $s_{\rm red}^{2}$ threshold from $0.001$ to $0.1$ and observed the number of archetypes associated with each galaxy subtype (ELGs, LRGs, and BGS) and the computational time. For a small $s_{\rm red}^{2}$ algorithm selects many similar archetypes as expected. We identify the optimal threshold parameter by noting when the number of archetypes for any subtype significantly changes to a very small number. In our methodology, we optimize the $s_{\rm red}^{2}$ value with respect to the number of LRG archetypes. This is a choice informed by the predominantly passive nature of LRGs, which results in fewer distinct spectral features compared to ELGs or BGS. We observe a significant reduction in the number of LRG archetypes, decreasing from $14$ to $2$, as the $s_{\rm red}^{2}$ increases from approximately $0.003$ to $s_{\rm red}^{2}\approx 0.005$. Moreover, the number of LRG archetypes is reduced to $0$ if we increase $s_{\rm red}^{2}$ to 0.1. We also note that the $1$ or $2$ archetypes cannot effectively model DESI LRG's observed features. \modify{We show the variation of the number of archetypes for each galaxy class as a function of $s^{2}_{\rm red}$ in Figure~\ref{fig:sred_square}.} Finally, we select the $s_{\rm red}^{2}\approx 0.003$ (somewhere between a very low to very high number of LRG archetypes) and generate our final set of galaxy archetypes. 

After running the \textit{SetCoverPy} with this choice, we get 385 (14 LRGs, 104 BGS, and 267 ELGs) archetypes from an input set of 1000 spectra. Note that there are more archetypes for ELGs than other subtypes. This is due to the substantial variations in the emission line properties of ELGs, which are usually associated with their color and stellar properties. In the top right panel of Figure~\ref{fig:archetypes}, we show the color ($g-r$) vs stellar mass for our final set of archetypes. One can see the archetypes represent the parent sample (left panel) distribution well in this plane, which shows that \textit{SetCoverPy} performs fairly well in solving the set covering problem for our purpose. Additionally, we also present representative spectra of ELGs (in blue), BGS (in orange), and LRGs (in red) in the bottom panel, showing distinctive spectral features characteristic of each galaxy sub-class. \tb{We also observe that there are few LRGs with $g-r>2$ in the parent galaxy sample while there are none in the archetype sample. The reason is possibly due to very limited distinct features intrinsic to LRGs that slightly differ only in their continuum amplitudes rather than shape. In fact, we see that a few ($ \sim 4$) LRG archetypes can actually model the majority of the parent LRG sample. We show those LRGs in Appendix~\ref{archetype_vs_parent}.}

Finally, we also compare some more parent and archetype galaxy properties in Appendix~\ref{archetype_vs_parent} to understand how well our archetypes span parent galaxies in those dimensions. We plan to explore ways to construct optimal samples of archetypes using a combination of galaxy properties and clustering methods (also see section~\ref{improvements}).

\subsection{Code Implementation}\label{code_details}

Redrock is written in \package{python} and updated and maintained as a part of the DESI public GitHub repository\footnote{\url{https://github.com/desihub/redrock}}. It is a high-performance parallel code that uses MPIs and multiprocessing to run on CPUs and GPUs. The code can fit thousands of spectra, classify their spectral type, and measure redshifts within a few minutes. This is crucial to analyzing the ongoing DESI survey as it \tb{will observe} more than $40$ million objects in the next five years. Our updated algorithm is now part of that repository, where we describe the newly added functional arguments in detail. The galaxy archetypes can be downloaded by following the instructions in the Redrock repository. 
 
The archetype code can also run on both CPUs and GPUs; however, the only remaining step that is not GPU accelerated yet is the \package{scipy's} \bvls method, which we plan to explore in the future. GPUs significantly improve the runtime for our archetype method, which we illustrate below. These spectral datasets are processed on NERSC's\footnote{National Energy Research Scientific Computing Center, \url{www.nersc.gov}} Perlmutter supercomputer GPU nodes, each equipped with $4$ GPU cores (each with $40$ GB memory) and $64$ CPU cores (with a total of 256 GB memory). For each of the tiles we analyzed, there were $10$ coadded spectra files (one for each DESI spectrograph), each containing $500$ targets (including real and sky fibers). Our archetype processing (including modeling in each camera) exhibited an average processing time of approximately $\sim 70$ seconds for $500$ spectra when utilizing GPU and CPU resources on NERSC's GPU nodes. We also note that \textit{BVLS} step (described in section~\ref{archetype_model}) is one of the slowest steps as it is an iterative method to estimate coefficients. In contrast, processing the same spectral files solely on CPU nodes (each having 64 cores with a total of 512 GB memory) required an average of $\sim 90$ seconds ($\sim 30$ \% slower than CPU+GPU runtime). For reference, when employing the current version ($0.19.0$) of Redrock (without archetypes), the processing time is $40\, \rm s$ ($\sim 2$ times faster than archetype) to analyze $500$ spectra on NERSC's CPU-only nodes, while the computation time reduces to $\sim 30\, \rm s$ ($\sim 3$ times faster) on NERSC's CPU+GPU nodes. This is expected, given that the Redrock (without archetypes) employs \package{numpy}'s (or \package{cupy}'s) linear algebra solvers to solve for PCA coefficients, which is significantly faster compared to the iterative BVLS method.

We emphasize that the new approach is significantly slower (by a factor of $\sim 2-3$) than the Redrock (without archetypes) implementation. However, it is still practical for analyzing large datasets containing millions of spectra. This trade-off in processing time allows for improved physical modeling of spectra and better redshift measurement, and spectral classification and holds significant promise for extending it to other large spectroscopic surveys in the future. 

\begin{figure*}
    \centering
     \includegraphics[width=0.975\linewidth]{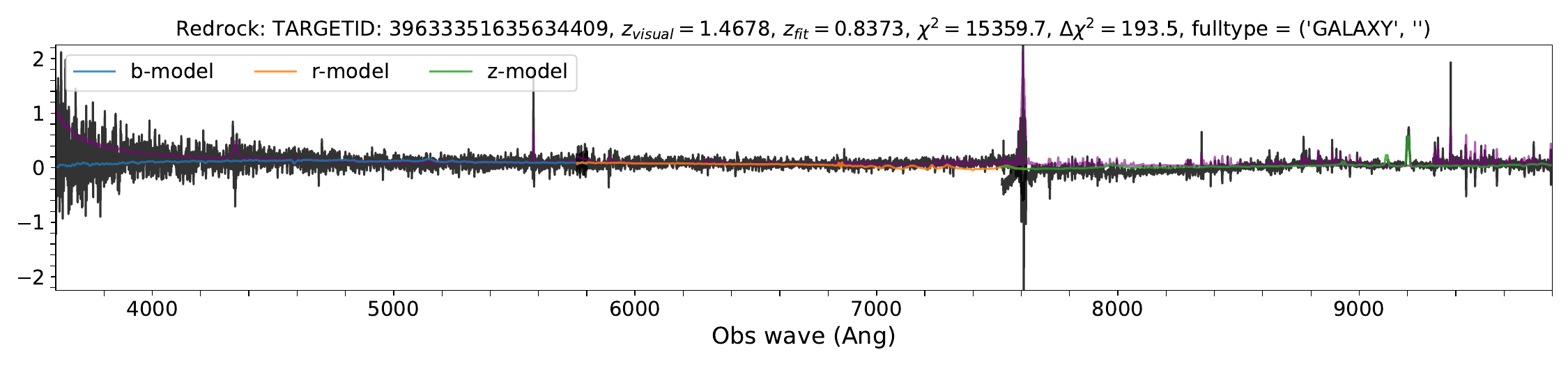}\\
     \includegraphics[width=0.975\linewidth]{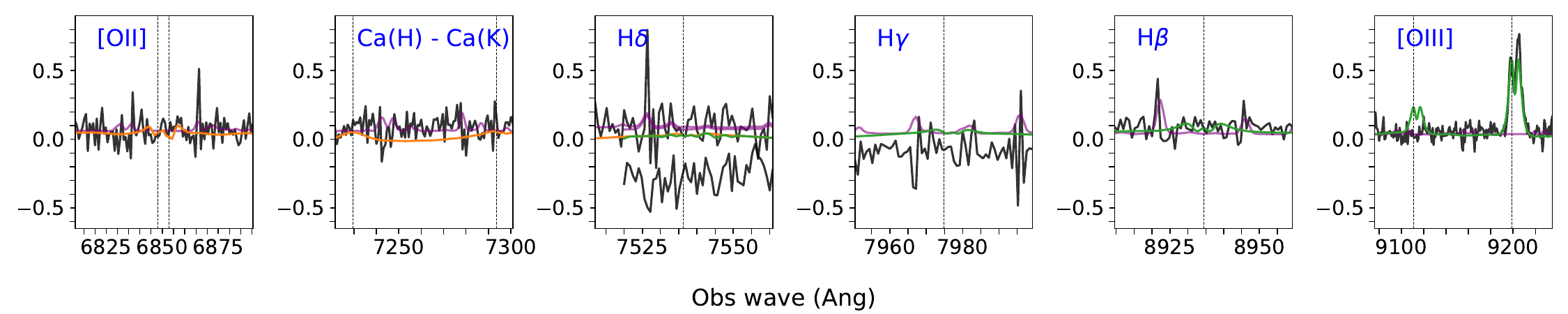}\\
    \includegraphics[width=0.975\linewidth]{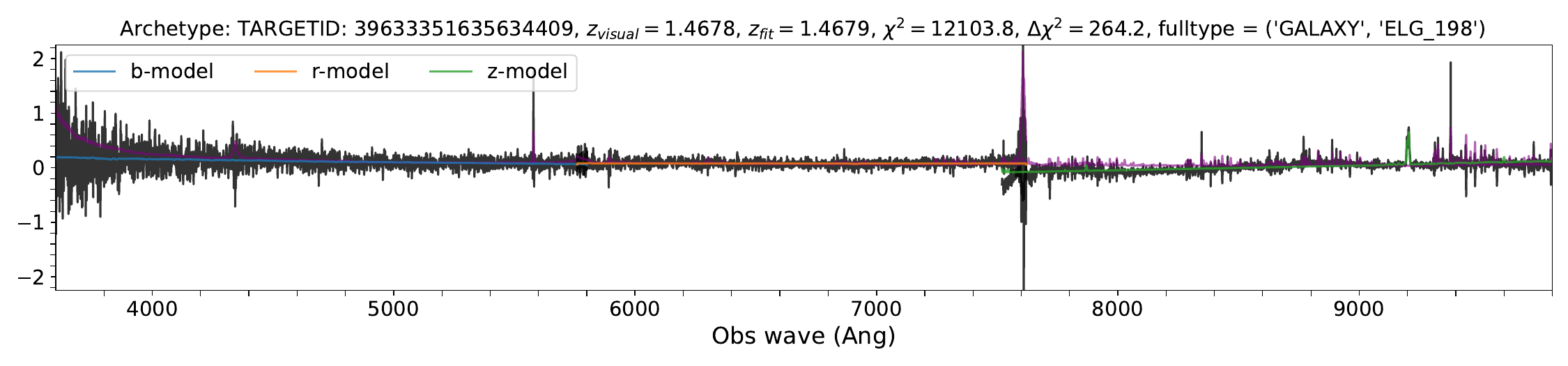}\\
	\includegraphics[width=0.975\linewidth]{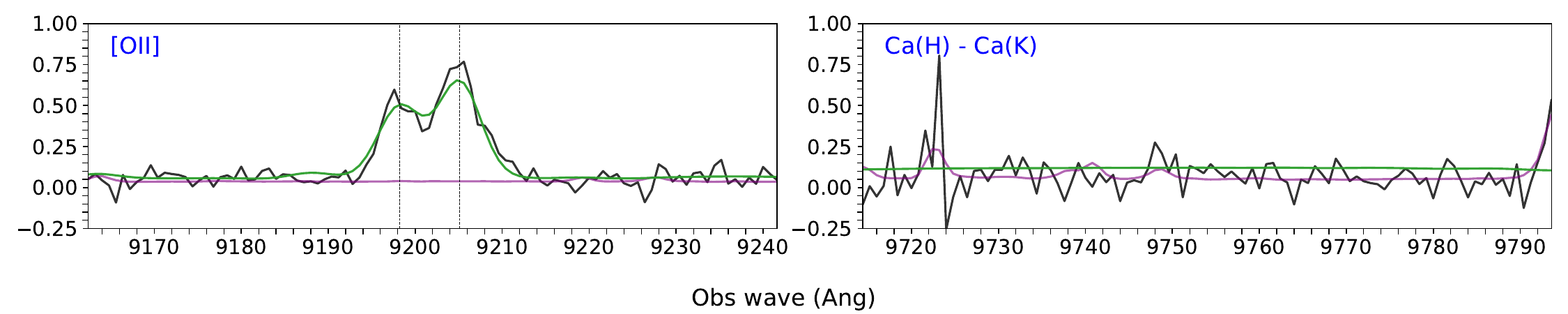}\\
    \caption{\textbf{First:} PCA-template-based best-fit model (shown in colored solid lines) of a galaxy spectrum. We can see the flux discontinuity in the $z$ camera ($\lambda \sim 7550$ \AA). \textbf{Second:} Zoomed version of PCA model of the same spectrum, showing the expected location of emission lines. \tb{The continuum is not well-fitted with PCA at the camera boundary. \oiii doublet is also not well fitted.} \textbf{Third:} Archetype-based per camera best-fit model of the same spectrum. The fit looks better visually and absorbs the flux discontinuity in the $z$ camera more accurately, also evident by smaller $\chi^{2}$. \textbf{Fourth:} Zoomed version of archetype-based per-camera model of the same spectrum, \tb{showing the expected location of emission lines.} \oii emission line ($\lambda \sim 9200$~\AA) is more accurately modeled. The black curve is the observed flux, the purple curve shows the error spectra in all panels, and the fitted model is shown in solid color lines.}
    \label{fig:spec1}
\end{figure*}

\begin{figure*}
    \centering
    \includegraphics[width=0.975\linewidth]{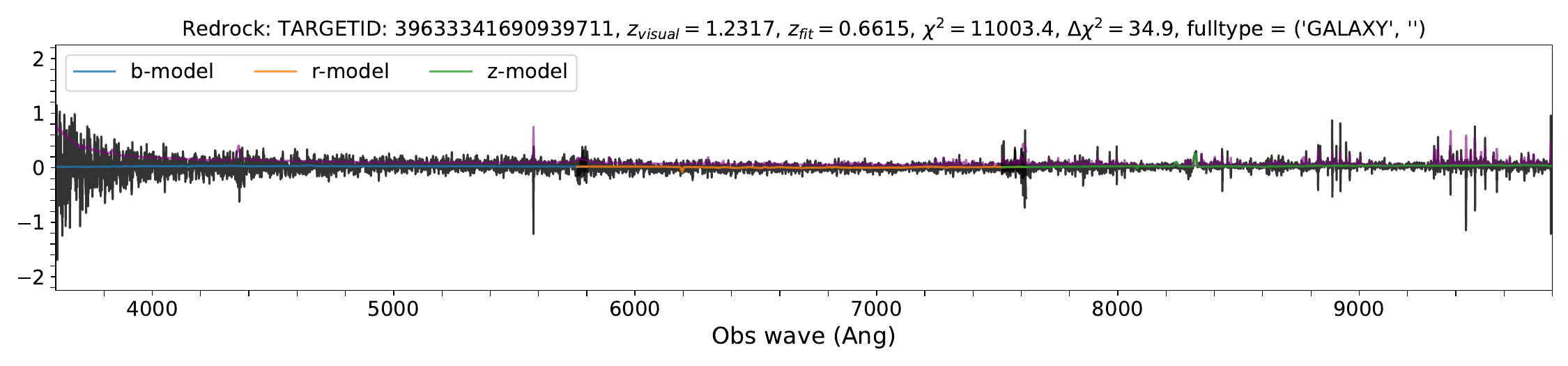}\\
     \includegraphics[width=0.975\linewidth]{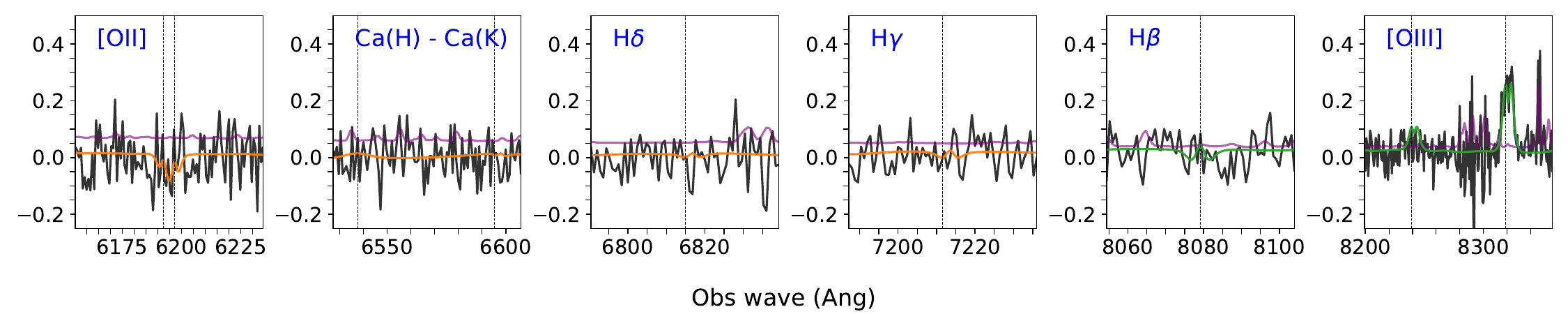}\\
    \includegraphics[width=0.975\linewidth]{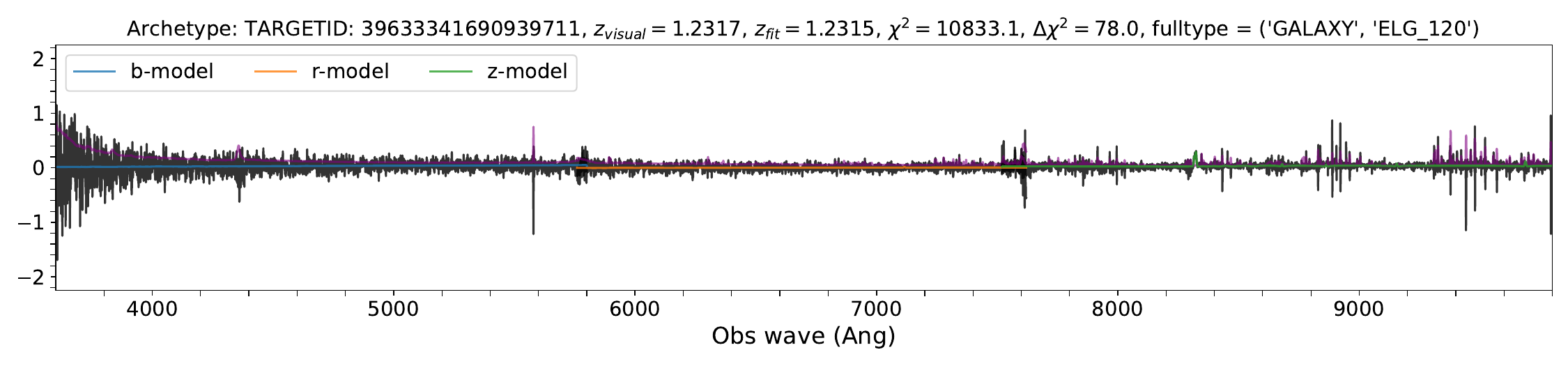}\\
	\includegraphics[width=0.975\linewidth]{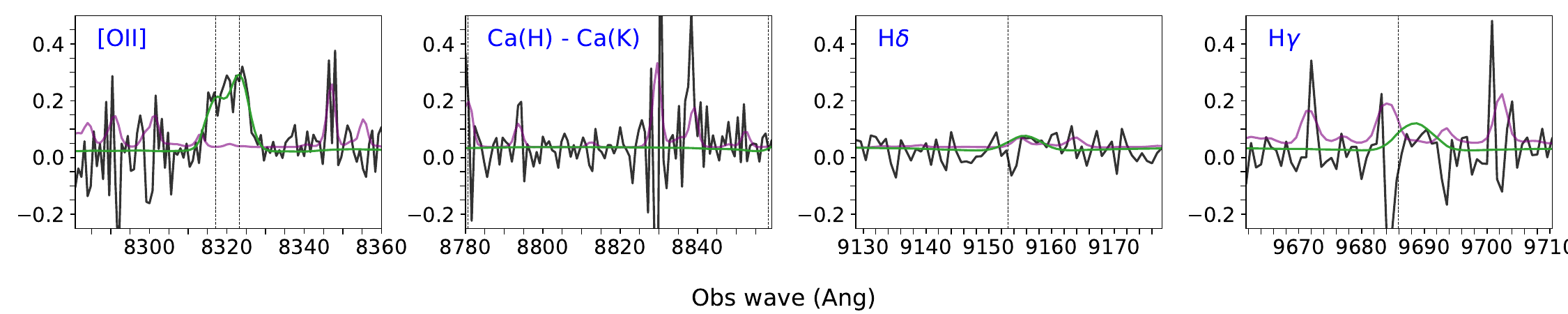}\\
    \caption{\textbf{First:} PCA-template-based best-fit model (shown in colored solid lines) of a galaxy spectrum.  \textbf{Second:} Zoomed version of PCA model of the same spectrum. We can see the unphysical negative absorption line fitting for \oii ($\lambda \sim 6200\,$\AA) and H$\beta$ ($\lambda \sim 8080\,$\AA) lines. In fact, Redrock(without archetypes) misidentifies \oii doublet as \oiii.
    \textbf{Third:} Archetype-based best-fit model of the same spectrum. \textbf{Fourth:} Zoomed version of archetype-based per-camera model of the same spectrum. The new model is more physical, and the redshift is more accurate, as confirmed by visual inspection of \oii doublet. The black curve is the observed flux, the purple curve shows the error spectra in both panels, and the model is shown in solid color lines.}
    \label{fig:spec2}
\end{figure*}

\section{Tests and Performance}\label{code_performance}

In this section, we illustrate the comprehensive tests conducted to assess the performance and efficiency of our archetype method applied to the DESI dataset. The analysis involved measuring the method's performance across various datasets characterized by distinct observing conditions, exposure times, signal-to-noise, and visual inspection criteria. We compare our test results against the existing DESI redshifts \tb{from PCA-only redrock, i.e., without archetypes (version \target{0.19.0})}. To provide a comprehensive comparative study, we run our archetype method on three distinct datasets: a) SV tiles that were observed on several nights to quantify product reproducibility of the pipeline, b) SV deep coadded spectra that were also visually inspected, and c) an extensive test run encompassing millions of targets selected from main survey data of Y1. For all the tests in the next sections, we use the nine best redshifts ($N_{\rm zbest}=9$, refer Figure~\ref{fig:method_flowchart}) of PCA-template-based redshift scan. \tb{We performed some tests exploring $N_{\rm zbest}$ (see section~\ref{more_tests}) where we changed it between $3$ to $15$ and found 9 to be the optimal value.} We find that this value is neither very small (which might miss the true redshift) nor very large (computationally expensive). We show all the comparison statistics for our archetype method with ($\sigma_{\rm a}=0.1$, Eqn~\ref{eqn:chi2_prior}) and without prior with redrock (without archetypes) in the following sections.

\begin{figure*}
    \centering
    \includegraphics[width=0.95\linewidth]{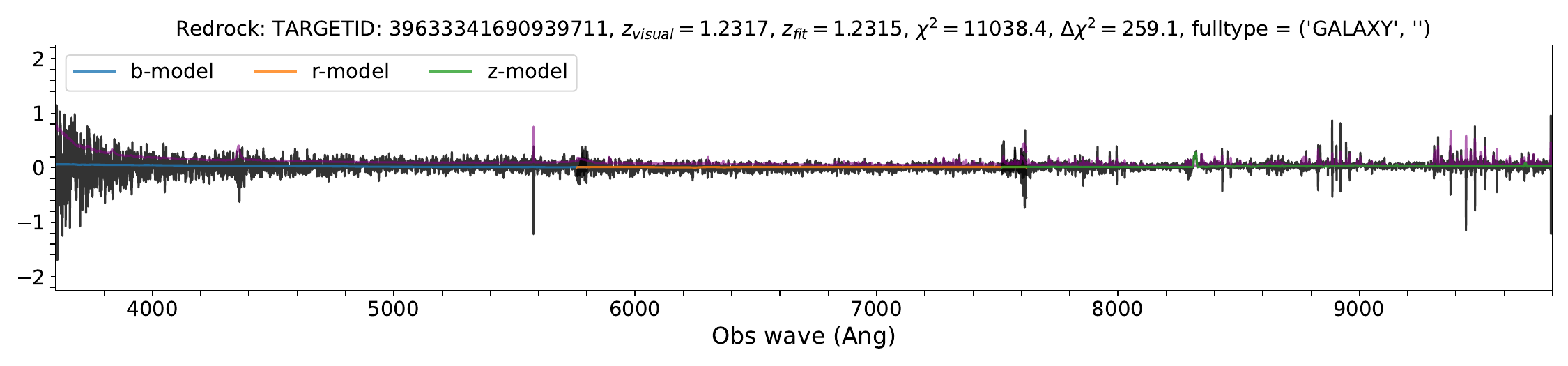}\\
	\includegraphics[width=0.95\linewidth]{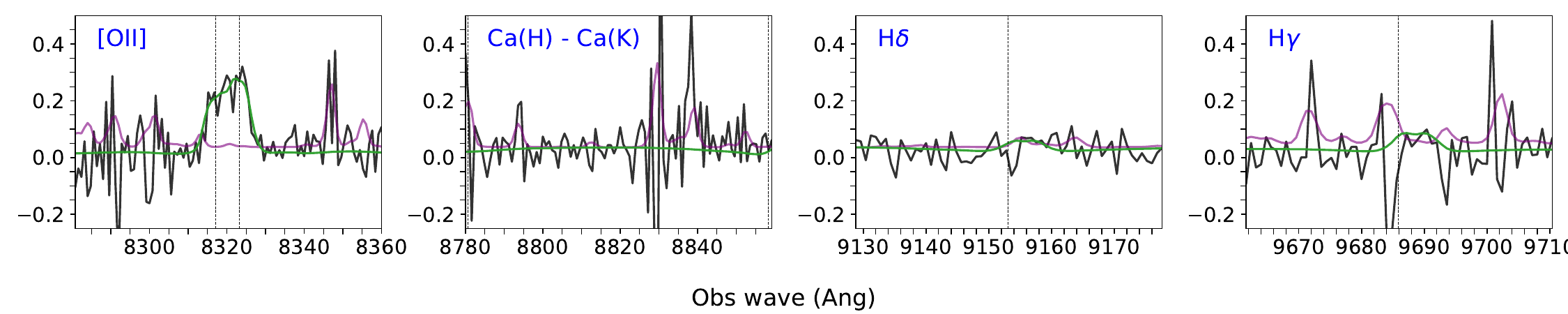}\\
    \includegraphics[width=0.95\linewidth]{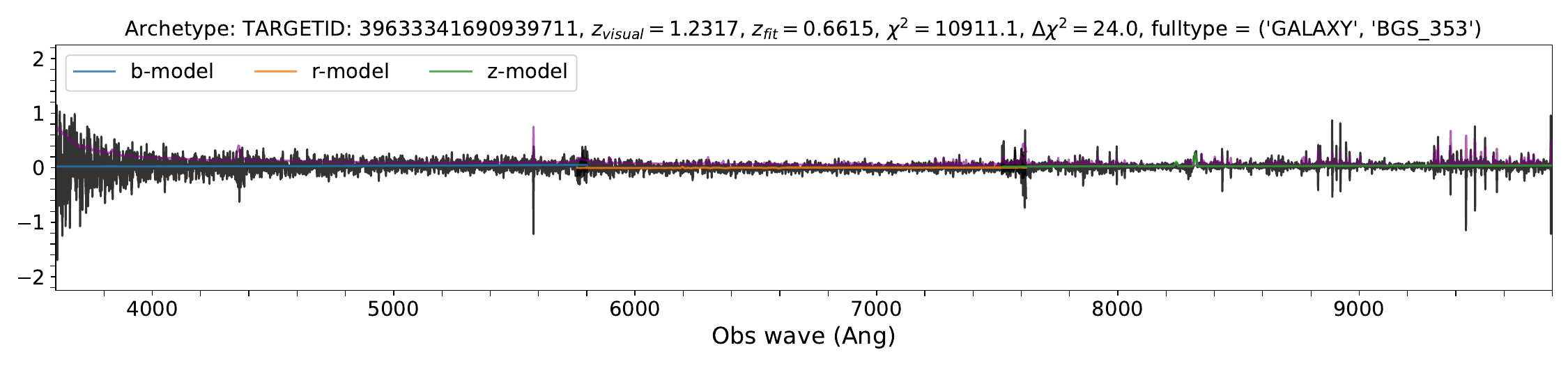}\\
	\includegraphics[width=0.95\linewidth]{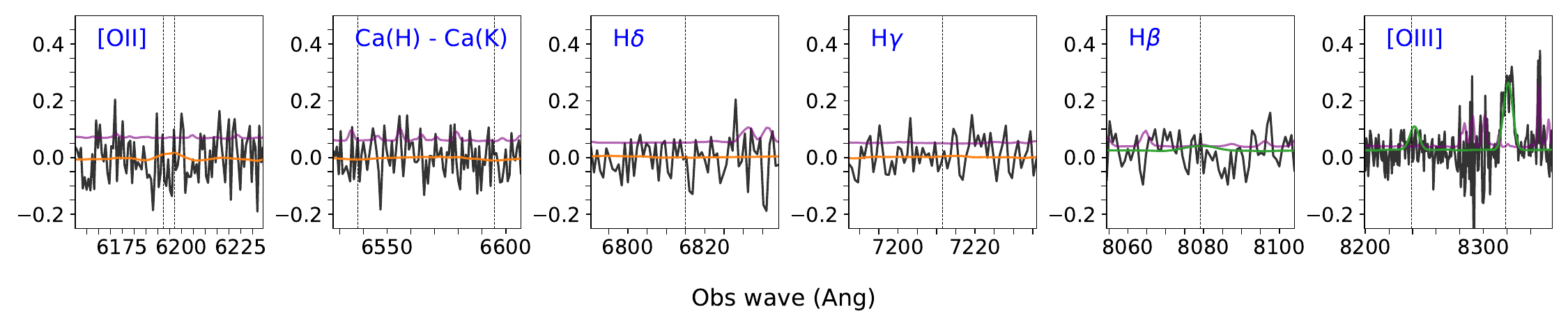}\\
    \caption{\textbf{First:} PCA-template-based model of the same galaxy spectrum (shown in Figure~\ref{fig:spec2}) at right redshift. \textbf{Second:} Zoomed version of the model. We can see that $\chi^{2}$ is higher than the best-fit model (see Figure~\ref{fig:spec2}).
    \textbf{Third:} Archetype-based fit of the same spectrum at wrong redshift. \textbf{Fourth:} Zoomed version of the archetype model. We see that our archetype model does not fit any unphysical feature, and $\chi^{2}$ is higher than the best-fit model (see third panel of Figure~\ref{fig:spec2}). The black curve is the observed flux, the purple curve shows the error spectra in both panels, and the model is shown in solid color lines.}
    \label{fig:pca_right_redshift}
\end{figure*}

\subsection{Archetype fitting of DESI spectra}\label{archetypefit}

As described in section~\ref{method}, our model is more physical than the PCA-based approach and can absorb the pipeline defects more accurately. In figures~\ref{fig:spec1} and ~\ref{fig:spec2}, we present two example spectra to illustrate this comparison between the redrock without archetypes and the archetype-based model. Figure~\ref{fig:spec1} shows a spectrum where the CCD discontinuity is clearly visible, and the spectrum has shifted vertically in $z-$camera ($\lambda>7500$~\AA). \tb{The spectrum was visually inspected and assigned a redshift of $1.4678$. However, the current PCA model (redrock without archetypes, first panel) fails to find the correct redshift. In fact, it misidentifies the \oiii doublet as \oii (see $\lambda\sim 9200$\,\AA). At the same time, our archetype approach (third panel) finds the correct redshift (see the nice \oii doublet at $\lambda\sim 9200$\,\AA), fits the data better, and absorbs the vertical CCD discontinuities in the spectrum more accurately.} This is evident not only by eye but also in the final $\chi^{2}$ values, where the archetype $\chi^{2}=12104$ is notably lower than the PCA-only (without archetypes) fit ($\chi^{2}=15360$). 

Next, in Figure~\ref{fig:spec2}, \tb{we demonstrate another example spectrum where redrock fails to find the correct redshift due to an unphysical negative emission line model of} \oii at $\lambda \sim 6200$\,\AA\, and $H\beta$ model at $\lambda\sim8080$ \AA\, (top panel). At the same time, our archetype model (third panel) shows improved physical modeling of the spectrum, yielding a more accurate redshift estimate. \tb{This is evidenced} by the location of \oii ($\lambda \sim 8320$\,\AA) line corresponding to the newly determined redshift, where we can see the emission features (also confirmed by small noise in the wavelength region). This spectrum was also visually inspected by DESI collaborators, who assigned a redshift of $1.2317$, and our archetype redshift estimate matches that. \tb{Notably, the $\chi^{2}$ is smaller while $\Delta \chi^{2}$ is larger in our archetype model than that of the PCA-only (without archetypes) run. The higher $\Delta \chi^{2}$ indicates} that the next best-fit model is far less probable. The wrong redshift estimate in the PCA-only (without archetypes) model is likely due to the flexibility of PCA to fit the negative feature as \oii line, which possibly drives the $\chi^{2}$ to be smaller than the model at the right redshift. To test this hypothesis in Figure~\ref{fig:pca_right_redshift}, we show the redrock model (without archetypes) at the right redshift ($z=1.2317$) and the archetype model at the wrong redshift ($z=0.6615$).

\tb{The top panel shows the PCA model for the spectrum estimated at the "right" redshift ($z=1.2317$). It is clear that despite the PCA model fitting the physical features of the spectrum well (see zoomed version), the estimated $\chi^{2}=11038$ is larger than the best-fit redshift ($\chi^{2}=11003$, compare top panel of Figure~\ref{fig:spec2}). This confirms our hypothesis that the PCA model is driven by fitting the unphysical negative dip at $\lambda\sim 6200$~\AA.\, In the bottom panel, we show the archetype model estimated at the "wrong" redshift ($z=0.6615$). As expected we find that $\chi^{2}=10911$ is larger than the best-fit $\chi^{2}=10833$ (bottom panel of Figure~\ref{fig:spec2}). Furthermore, this model does not fit the unphysical negative dip (see zoomed version around $\lambda\sim 6200$\,\AA), even at the wrong redshift. This is because our archetypes are physical galaxy models, and their coefficients are always positive in our model, which naturally prohibits any such unphysical fitting.} 

\begin{table*}
\centering
  \caption{Survey validation repeat tiles redshift success comparison- Redrock (without archetypes) vs. archetype method. $X$  and $X_{\rm o}$ are the \textit{good redshift criteria} parameters for the corresponding target classes with nominal effective exposure time as described in Table~\ref{tab:z_success}. The catastrophic failure and redshift purity are defined in the text. We also apply the following quality cuts ($Q_{\rm o}$) on single epoch data. \textbf{a) BGS}: $\rm ZWARN=0,\, X>X_{\rm o}$, \textbf{b) LRG}: $\rm ZWARN=0$, $z_{\rm pernight}<1.5, \, X>X_{\rm o}$, \textbf{c) ELG}: $X>X_{\rm o}$ and $\rm [OII]_{flux},\,\sigma_{[OII]_{flux}}>0$, i.e. valid measurements of single epoch \oii fluxes and their errors. $z_{\rm pernight}$ and \target{ZWARN} denote the redshift and its warning bits obtained for single epoch spectra.}
  % prior values are not the latest ones
  \begin{tabular}{ccccccccccc}
    \hline
    Target&$N_{\rm target}$&\multicolumn{3}{c}{$N$ (selected by quality cut, $Q_{\rm o}$)}&\multicolumn{3}{c}{$N (Q_{\rm o},\,\,|\Delta v|> 1000$ \kms)}&\multicolumn{3}{c}{Redshift Purity (\%)} \\
    class&&&&&\multicolumn{3}{c}{\textit{Catastrophic failure}}&\multicolumn{3}{c}{(defined in the text)} \\
    \hline
    &&Redrock & Archetype  & Archetype& Redrock & Archetype  & Archetype&Redrock & Archetype  & Archetype \\
    &&  &(no prior)&($\sigma_{\rm a} =0.1$)&  &(no prior)&($\sigma_{\rm a} =0.1$)& &(no prior)&($\sigma_{\rm a} =0.1$)\\
    \target{BGS\_FAINT}&2267&2178&2183&2184&9&8&8&99.58&99.63&99.63\\
    \target{BGS\_BRIGHT}&3256&3178&3184&3187&11&13&11&99.65&99.60&99.66\\
    \target{LRG}&9981&9556&9625&9633&80&62&62&99.16&99.36&99.36\\
    \target{ELG\_LOP}&9806&7647&7658&7647&31&25&19&99.59&99.67&99.75\\
    \target{ELG\_VLO}&2417&2327&2326&2332&6&2&3&99.74&99.91&99.88\\
    \target{ELG\_HIP}&1136&944&953&951&1&1&2&99.89&99.89&99.81\\

    \hline
  \end{tabular}
  \label{tab:pernight_comparison}
\end{table*}

\subsection{Survey Validation Repeat Observations}\label{repeat_observations}

\begin{figure*}
    \centering
    \includegraphics[width=0.475\linewidth]{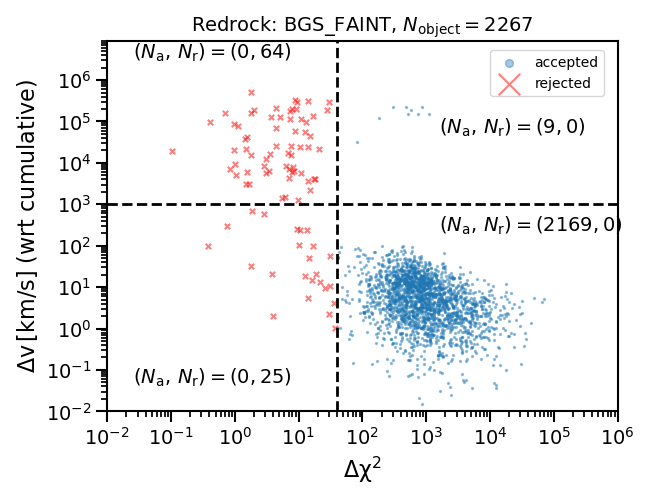}
	\includegraphics[width=0.475\linewidth]{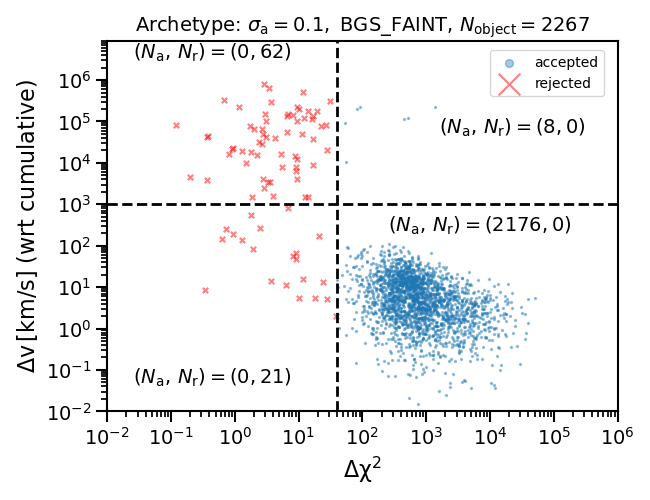}\\
	\includegraphics[width=0.475\linewidth]{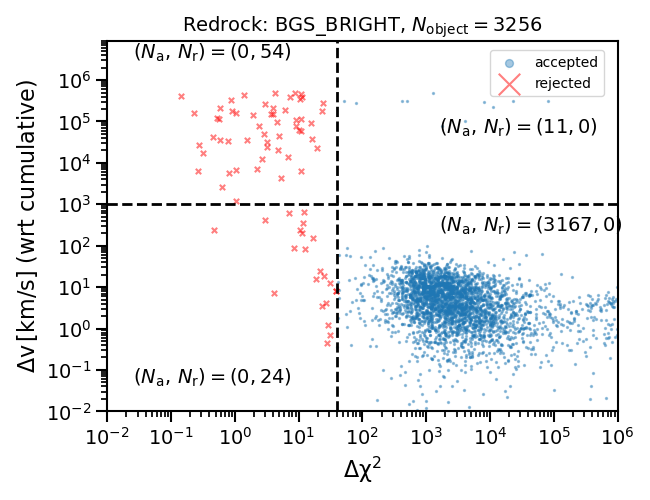}
	\includegraphics[width=0.475\linewidth]{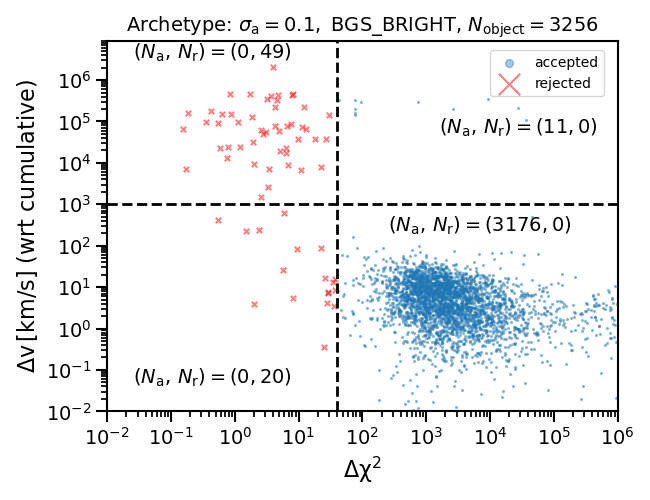}\\
    \caption{The redshift difference between single exposure and deep coadded spectra of same \target{BGS\_FAINT} (top) and \target{BGS\_BRIGHT} (bottom) targets. The $\Delta \chi^{2}$ values are taken from single epoch observations. The vertical dashed line ($\Delta \chi^{2}>40$) shows the boundary to \tb{define confident redshift}, and the horizontal dashed line defines the catastrophic redshift failure, i.e., $|\Delta v|>1000$ \kms. The left panel shows the results for Redrock (without archetype), and the right panel shows the results for our archetype approach (with a prior of $\sigma_{\rm a}=0.1$ on polynomial coefficients). \tb{In both panels, the blue dots show the accepted redshifts ($N_{\rm a}$) while the red crosses show the rejected redshifts ($N_{\rm r}$).} To guide the readers, the bottom right quadrant in each subpanel represents good performance, showing that individual night redshifts match with deep coadded redshifts of the same spectrum, while the upper right shows the redshifts that differ from the deep coadded spectrum and also flagged as very robust redshift in the pipeline.}
    \label{fig:bgs_bright_repeat}
\end{figure*}

\begin{figure*}
    \centering
	\includegraphics[width=0.49\linewidth]{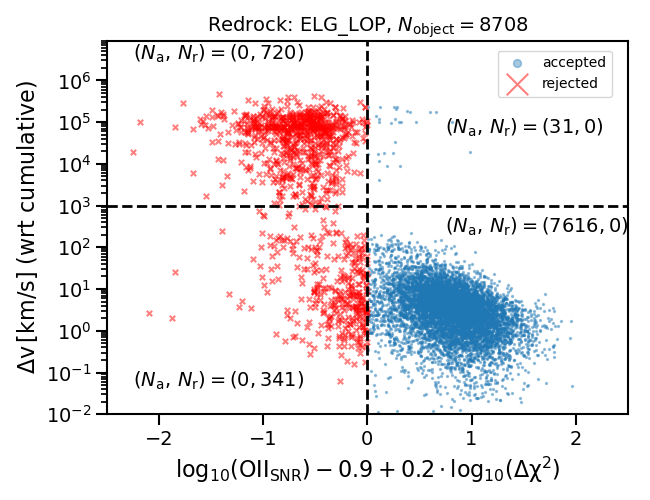}
	\includegraphics[width=0.49\linewidth]{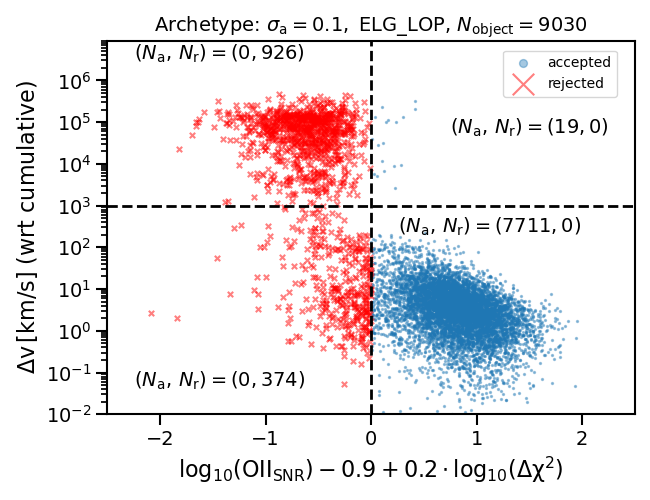}\\
    \includegraphics[width=0.49\linewidth]{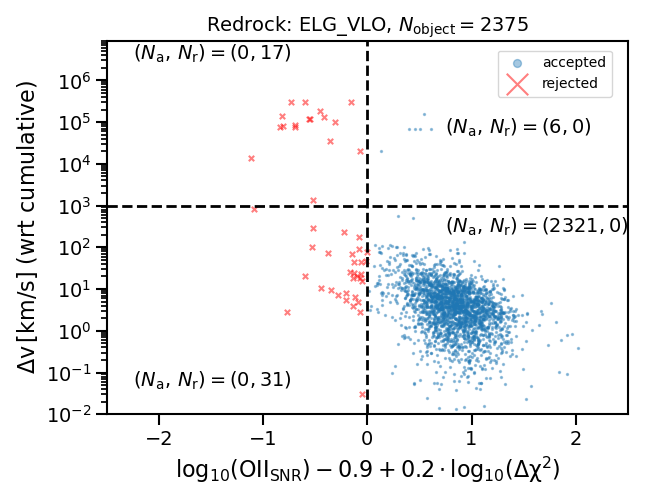}
	\includegraphics[width=0.49\linewidth]{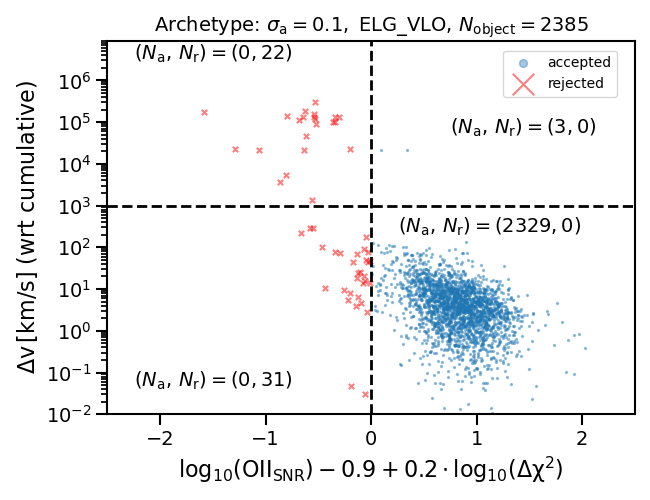}\\
    \caption{The redshift difference between single epoch observation and deep coadded spectra of same \target{ELG\_LOP} targets. The good redshift selection criteria are defined in terms of \oii SNR and $\Delta \chi^{2}$ where both quantities are taken from single epoch observations (see Table~\ref{tab:z_success}). We only include spectra that have valid \oii line strength (i.e., $\rm [O II]_{\rm flux}>0$ and $\rm \sigma_{[O II]_{\rm flux}}>0$). The vertical dashed line ($X>0$, see Table~\ref{tab:z_success}) shows the boundary to define good redshift, and the horizontal dashed line defines the catastrophic redshift failure, i.e., $|\Delta v|>1000$ \kms. The left panel shows the results for Redrock (without archetypes), and the right panel shows the results for our archetype approach (with a prior of $\sigma_{\rm a}=0.1$ on polynomial coefficients). \tb{In both panels, the blue dots show the accepted redshifts ($N_{\rm a}$) while the red crosses show the rejected redshifts ($N_{\rm r}$).} To guide the readers, the bottom right quadrant in each subpanel represents good performance, showing that individual night redshifts match with deep coadded redshifts of the same spectrum, while the upper right shows the redshifts that differ from the deep coadded spectrum and also flagged as very robust redshift in the pipeline. Archetype performs $\sim50$ \% better in reducing catastrophic redshift failure.}
     \label{fig:elg_repeat}
\end{figure*}

\begin{figure*}
    \centering
	\includegraphics[width=0.475\linewidth]{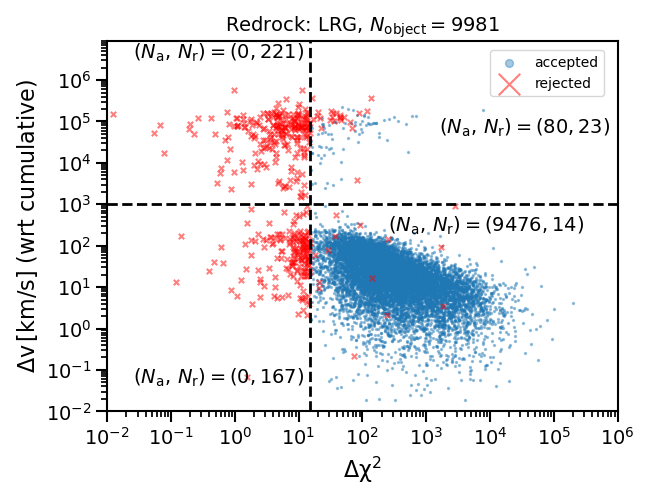}
	\includegraphics[width=0.475\linewidth]{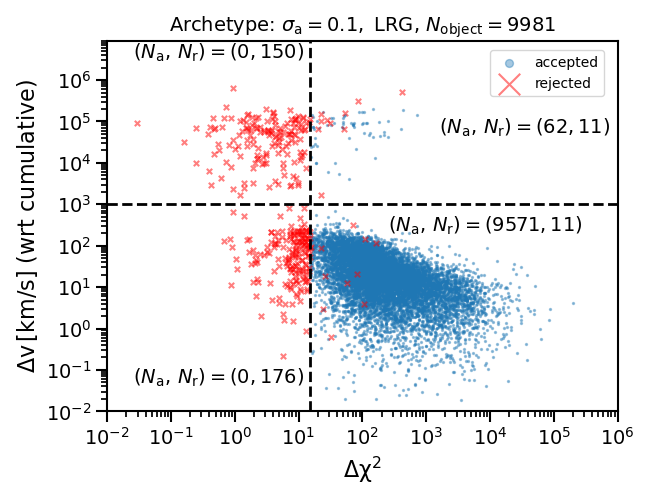}\\
    \caption{The redshift difference between single epoch and deep coadded spectra of the same LRG targets. The $\Delta \chi^{2}$ values are taken from single epoch observations. The vertical dashed line ($\Delta \chi^{2}>15$) shows the boundary to define good redshift, and the horizontal dashed line defines the catastrophic redshift failure, i.e., $|\Delta v|>1000$ \kms. The left panel shows the results for Redrock (without archetypes), and the right panel shows the results for our archetype approach (with a prior of $\sigma_{\rm a}=0.1$ on polynomial coefficients). \tb{In both panels, the blue dots show the accepted redshifts ($N_{\rm a}$) while the red crosses show the rejected redshifts ($N_{\rm r}$).} To guide the readers, the bottom right quadrant in each subpanel represents good performance, showing that individual night redshifts match with deep coadded redshifts of the same spectrum, while the upper right shows the redshifts that differ from the deep coadded spectrum and also flagged as very robust redshift in the pipeline. \tb{There are some rejected redshifts in the bottom right quadrant. That is because we also reject LRGs with $z>1.5$.} Archetype performs $\sim20$ \% better in reducing catastrophic redshift failure.}
     \label{fig:lrg_repeat}
\end{figure*}

A total of six unique "tiles" were specifically designed to target distinct galaxy target classes in the SV phase. Three of these tiles were designed to observe ELGs (TILEIDs 80606, 80608, and 80610; see \citealt{raichoor2023ELG} for more details). As discussed earlier, ELGs were divided into \target{ELG\_HIP}, \target{ELG\_LOP}, and \target{ELG\_VLO}. Two additional tiles (TILEIDs 80605 and 80609, see \citealt{zhou2023LRG}) were designed to target LRGs, and one tile (TILEID 80613, see \citealt{hahn2023BGS}) was designed for BGS. As described in section~\ref{desispectra}, the BGS targets were further subdivided into \target{BGS\_BRIGHT} and \target{BGS\_FAINT}. 

These tiles were systematically observed in the SV phase on different nights between December 2020 and May 2021. The nominal exposure times (same for the main survey) varied depending on the target class, ranging from approximately $\sim180$ seconds for BGS targets to around $\sim 1000$s for LRG and ELG targets. In total, $8910$ unique galaxy targets were observed on these tiles, with $1153$ spectra attributed to \target{BGS\_BRIGHT}, $792$ to \target{BGS\_FAINT}, $2522$ to \target{ELG\_LOP}, $644$ to \target{ELG\_VLO}, $300$ to \target{ELG\_HIP} and $3499$ to \target{LRGs}. The individual exposures of these targets were observed multiple times, ranging from four to ten nights per target class spread across several nights. We ran the redshift fitter for these targets and compared their redshift estimates with those obtained from their deep coadded spectra (discussed in section~\ref{visual_tiles}) using redrock (without archetypes) and archetype-based per-camera approach and compared their performance. 

This test is crucial to quantify the performance of redshift fitter software as it shows its ability to analyze the spectra with varying observing conditions and S/N; therefore, understanding the software performance on single-epoch spectra is important. In addition, the test also allows us to empirically define `good redshift' criteria (defined in Table~\ref{tab:z_success}) for different target classes based on a combination of the rate of `catastrophic failure' and $\Delta \chi^{2}$ of the best-fit model. The catastrophic redshift failure is defined as the difference between redshift ($z_{\rm epoch}$) measured with low S/N single-epoch spectrum and redshift ($z_{\rm coadd}$) measured with deep coadded high S/N spectrum of the same target. If $|\Delta v| = \Big |\frac{z_{\rm epoch}-z_{\rm coadd}}{1+z_{\rm coadd}} \Big | \cdot c> 1000\,\, \rm km/s$, then the measurement is termed as catastrophic failure \citep{desisv2023}.

The detailed comparison statistics of redshift success and catastrophic failure can be found in Table~\ref{tab:pernight_comparison}. We present the redshift differences (expressed as $\Delta v$) between single-epoch observations and the deep coadds for the same DESI targets, as a function of the $\Delta \chi^{2}$ values obtained from single-epoch observations for \target{BGS\_FAINT}, \target{BGS\_BRIGHT}, \target{ELG}, \target{ELG\_LOP}, \target{ELG\_VLO}, and \target{LRGs} for PCA-only (without archetypes) model (left panels) and our archetype model (right panels) in Figures~\ref{fig:bgs_bright_repeat},~\ref{fig:elg_repeat} and \ref{fig:lrg_repeat}, respectively. To guide the readers, we define the meaning of each quadrant for these figures: bottom right is good, meaning individual nights agree with the deep coadd, and they have a highly confident quality metric; bottom left are missed opportunities: they agreed with deep coadd but weren't flagged as confident, upper left are failures, but at least they weren't flagged as confident, upper right are the worst: catastrophic failures with redshift discrepancies but flagged as confident. \tb{In all figures, the blue dots refer to the accepted redshifts ($N_{\rm a}$), while red crosses refer to the rejected redshifts ($N_{\rm r}$). Note that the sum of blue and red points in all samples is equal to the number of targets shown in Table~\ref{tab:pernight_comparison}, except for ELG targets, where the sum of plotted points is smaller than the total number of targets in the table. This is because we select spectra having positive \oii flux and their errors before we calculate \oii SNR that is plotted in Figure~\ref{fig:elg_repeat}.}

\modify{Next, we define additional statistics for these datasets. Utilizing the truth tables for redshifts (i.e., the visual redshifts) of these targets, we calculate the precision, recall, and F1 scores for both methods. The results are presented in Table~\ref{tab:recall_precison} in Appendix~\ref{more_statistics}. Overall, these metrics are very high for both methods, indicating their excellent performance on the DESI datasets. However, the metrics are marginally higher for the archetype method (with $\sigma_{\rm a}=0.1$). Here, we also point out that we can not perform such analysis for main survey targets, as there are no truth tables for them. However, we expect very similar metrics for them as well, as many of these targets were also selected for the main survey and have comparable exposure times.}

We observe that both redrock (without archetypes) and archetype models (with prior $\sigma_{\rm a}=0.1$) perform equally well for almost all target classes. We find that the overall catastrophic redshift failure rate is relatively very small ($\gtrsim 0.5-1$\% of the full sample) in both models. \tb{This shows that the current PCA-based Redrock is already very efficient.} For the \target{BGS\_BRIGHT} targets (Figure~\ref{fig:bgs_bright_repeat}), our archetype-based approach (with $\sigma_{\rm a}=0.1$) yields $3187$ accurate redshifts ($\Delta \chi^{2}>40$) compared to $3178$ in the Redrock (without archetypes) run, while similar success is seen for \target{BGS\_FAINT} target classes ($2178$ vs. $2184$ in redrock (without archetypes) vs archetype with $\sigma_{\rm a}=0.1$), though slightly less catastrophic failure in archetype mode ($8$ vs $9$ in Redrock (without archetypes)). For \target{ELG\_LOP}, \target{ELG\_VLO} and \target{LRGs}, we see that in the archetype-based method, the relative improvement is $\sim 10-40$\% in reducing catastrophic redshift failure ($|\Delta v|>1000$~\kms, compare the blue dots in the upper right quadrant in each panel). For \target{ELG\_LOP} targets, the archetype-based approach (with $\sigma_{\rm a}=0.1$) yields $19$ catastrophic failures ($25$ without prior case), a $\sim 40$\% improvement compared to the $31$ failures in the Redrock (without archetypes) run (see Table~\ref{tab:pernight_comparison} and top panel of Figure~\ref{fig:elg_repeat}, for detailed comparison). Next, for \target{ELG\_VLO}, there are $3$ catastrophic failures in the archetype approach (with $\sigma_{\rm a}=0.1$) as opposed to $6$ in Redrock (without archetypes), showing a significant improvement (see bottom panel of Figure~\ref{fig:elg_repeat}). The performance is similar for both approaches for \target{ELG\_HIP}. Similarly, for LRGs, the archetype method delivers \tb{$\sim 25$\% (62 compared to 80 in Redrock (without archetypes))} less catastrophic redshift failures than redrock (see Figure~\ref{fig:lrg_repeat}). Also, the redshift success is slightly lower in Redrock (without archetype) mode ($9556$ vs. $9625$). Furthermore, our archetype-based modeling approach consistently delivers marginally higher redshift purity, i.e., the fraction of accurate redshifts that satisfy redshift success criteria, $\frac{N (Q_{\rm o},\,\,|\Delta v| < 1000 \,\rm kms^{-1})}{N(Q_{\rm o})}$. $Q_{\rm o}$ is the \textit{good redshift criteria} in terms of $\Delta \chi^{2}$ and \oii$\rm_{SNR}$ for the corresponding target classes as described in Table~\ref{tab:z_success} and~\ref{tab:pernight_comparison}. The redshift purity for our archetype method is slightly higher ($0.1-0.2\%$) than the current Redrock. The relatively higher purity and lower catastrophic failures in the archetype approach underscore its improved performance in achieving precise redshift measurements. However, seeing more significant changes in these statistics would be our goal in the future.

\modify{It is important to emphasize that the overall performance of both Redrock (without archetypes) and archetype approach on DESI data well exceeds the quality thresholds necessary for cosmological analysis. For instance, the rates of catastrophic failure are significantly lower ($<1\%$) than the acceptable limit of $<5\%$ \citep{desisv2023}. Furthermore, the typical redshift errors achieved with the DESI pipeline meet the precision requirements for large-scale structure analysis. Specifically, the typical redshift errors for the BGS, ELGs, and LRGs are approximately $10$, $20$, and $40$ \kms, respectively, which are well below the required redshift precision of approximately 200 \kms \citep{desisv2023, lan2023}.}

\subsection{Survey Validation Visually Inspected Deep Tiles}\label{visual_tiles}

In this section, we describe the efficiency of the archetype method in measuring the redshifts for targets that DESI collaborators visually inspected during the SV phase. As described earlier, the single-epoch observations of these tiles were combined across exposures to construct higher S/N coadded spectra. The cumulative effective exposure times vary from $\sim1500$ seconds to around $\sim5000$ seconds. These spectra were visually inspected by at least two DESI collaborators, following a well-defined, uniform approach, which is detailed in \citet{lan2023}. In summary, visual inspectors look for the most prominent emission lines (e.g., $\rm H\alpha$, $\rm H\beta$, \oii, \oiii), 4000~\AA~break, and the shape of the continuum to verify the redshift and spectral class of DESI spectra estimated by redshift fitter. A set of predefined metrics, as described in \citet{lan2023}, is employed to maintain consistency in the inspection process. These metrics aid in assigning quality labels (QA\_VI)\footnote{The members assign a quality index between $0$ (not a robust redshift) to $4$ (most robust redshift) depending upon the spectral features visible in the spectra.} to each coadded spectrum based on spectral features to ascertain the reliability of the redshift estimates and spectral classes. We ran our archetype method on these targets and then compared our redshifts obtained from the pipeline and those determined through visual inspection. 

\begin{table}
\centering
  \caption{Visually inspected deep coadded tiles redshift comparison for redrock vs. archetype. Catastrophic redshift failure is defined as, $|\Delta v| = \frac{|z_{\rm ref} - z_{\rm visual}|}{1+z_{\rm visual}}\cdot c$, where $z_{\rm ref}$ and $z_{\rm visual}$ are pipeline and visual redshifts, respectively. We select spectra that are visually classified as ``good", i.e., $\rm QA\_VI\geq 2.5$ \citep[see][for details]{lan2023}.}
  \addtolength{\tabcolsep}{-0.4em}
  \begin{tabular}{ccccc}
    \hline
    Target&$N_{\rm target}$&\multicolumn{3}{c}{$|\Delta v|> 1000$ \kms} \\
    class&$(\rm QA\_VI \geq 2.5)$&\multicolumn{3}{c}{(Catastrophic failure)} \\
    
    \hline
    & &Redrock & Archetype  & Archetype \\
    & & &(no prior)&($\sigma_{\rm a} =0.1$)\\
    \target{BGS\_FAINT}&754&9&6&6\\
    \target{BGS\_BRIGHT}&1091&5&3&3\\
    \target{ELG\_LOP}&2132&25&21&21\\
    \target{ELG\_VLO}&639&2&1&1\\
    \target{ELG\_HIP}&263&2&2&2\\
    \target{LRG}&3440&39&26&26\\
    
    \hline
  \end{tabular}
  \label{tab:visual_comparison}
\end{table}

\begin{figure*}
    \centering
	\includegraphics[width=0.95\linewidth]{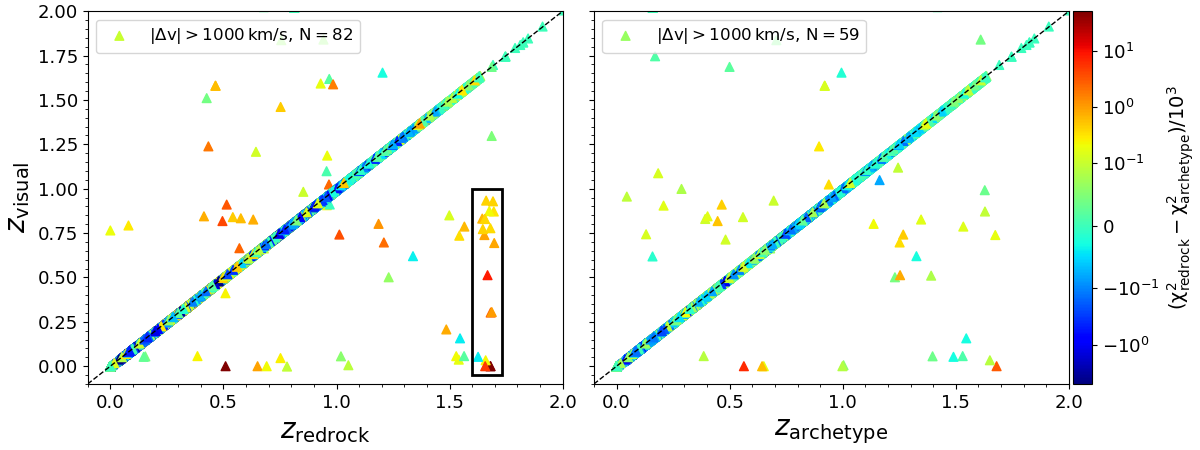}
    \caption{\modify{The redshift comparison between the PCA-based Redrock method (left) and the archetype method (right), color-coded by their $\chi^{2}$ difference for visually inspected galaxy spectra, reveals notable differences. The archetype method demonstrates reduced spurious clustering of redshifts and fewer catastrophic failures around $z\sim1, 1.6$. Furthermore, the archetype method generally yields lower $\chi^{2}$ values, indicating better model fits.}}
    \label{fig:z_comp_chi2}
\end{figure*}

The comparison results are presented in Table~\ref{tab:visual_comparison}, where we compare our results (with and without priors on Legendre terms) with redrock (without archetypes) measurements. Our approach shows a $10-30$\% improvement in reducing catastrophic redshift failure for all target classes. For example, in the case of \target{BGS\_BRIGHT}, the number of catastrophic redshift failures is $9$ in redrock (without archetypes) compared to $6$ in archetype (with prior $\sigma_{\rm a}=0.1$) case. Similarly, \target{ELG\_LOP} targets have $21$ catastrophic failure in the archetype (with prior $\sigma_{\rm a}=0.1$) mode, compared to $25$ in redrock (without archetype) mode. Finally, for \target{LRG}, there are only $26$ redshift failures as opposed to $39$ in redrock (without archetype). \modify{Next, we also compare the redshift estimates, color-coded by the $\chi^{2}$ difference between the PCA-only model (i.e., Redrock without archetypes) vs archetype model (with $\sigma_{\rm a}=0.1$) in Figure~\ref{fig:z_comp_chi2}. We see a spurious clustering at $z_{\rm redrock}\sim 1, 1.6$ in the redrock method (left), which goes away in the archetype method (right). The archetype method consistently yields lower $\chi^{2}$, indicating better model-fits for the input spectra.} These results clearly show that our method performs better than the existing method on deep coadded spectra.

\subsection{Large Test Runs on the Y1 dataset}\label{large_test}

As described above, our archetype approach has improved overall redshift estimates for targets observed with SV tiles. However, these are very small datasets (a few thousand only) relative to what DESI will provide in the coming years. Subsequently, a comprehensive large-scale test run was conducted on a large suite of tiles extracted from the Y1 dataset of DESI. Our test dataset included observations from 36 nights spread across May 2021 to June 2022, averaging 3 nights per month, encompassing a diverse range of observing and instrumental conditions. Within this dataset, 553 tiles were selected, comprising 233 dark and 320 bright tiles. This subset of data encompassed a total of $2,281,969$ target spectra and $339,712$ sky spectra selected from fibers pointing to targets and blank sky region\footnote{Each DESI coadded fits file includes a header storing \target{OBJTYPE} of observed TARGETIDs. A reader can apply the conditions of \target{OBJTYPE==TGT} and \target{OBJTYPE==SKY} to select target and sky fibers}, respectively. We used the archetype-based per-camera spectral fitting method and compared redshifts obtained with redrock (without archetypes).

\subsubsection{Performance on sky fibers}\label{zsky}

\begin{figure*}
    \centering
	\includegraphics[width=0.875\linewidth]{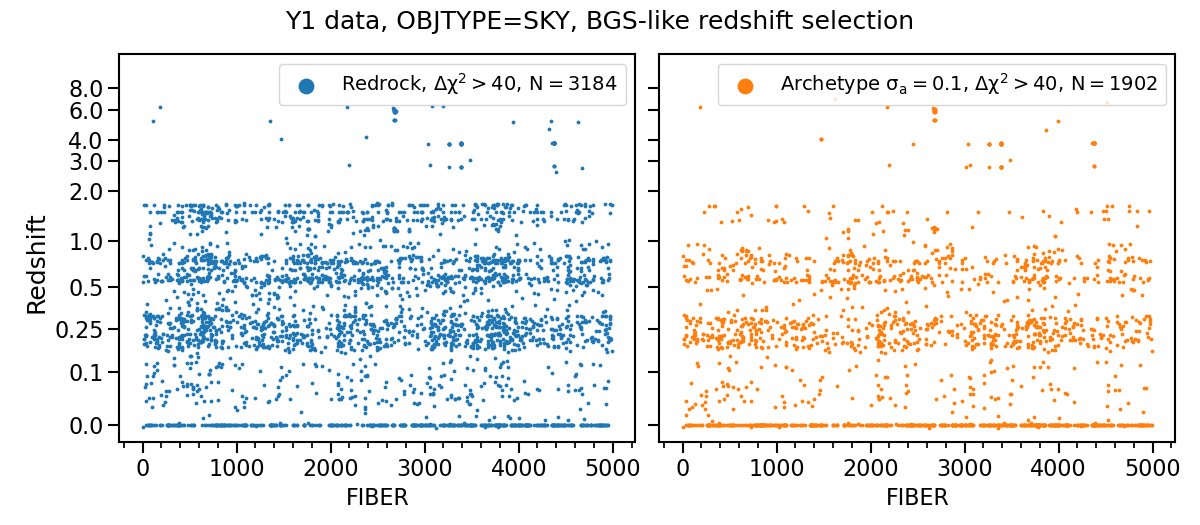}\\
	\includegraphics[width=0.875\linewidth]{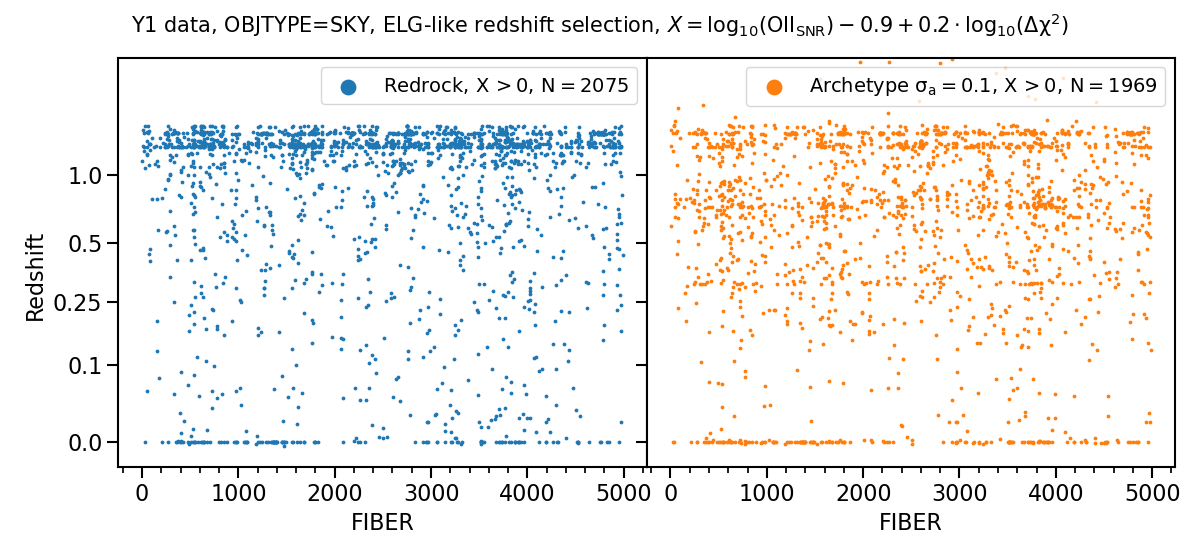}\\
	\includegraphics[width=0.875\linewidth]{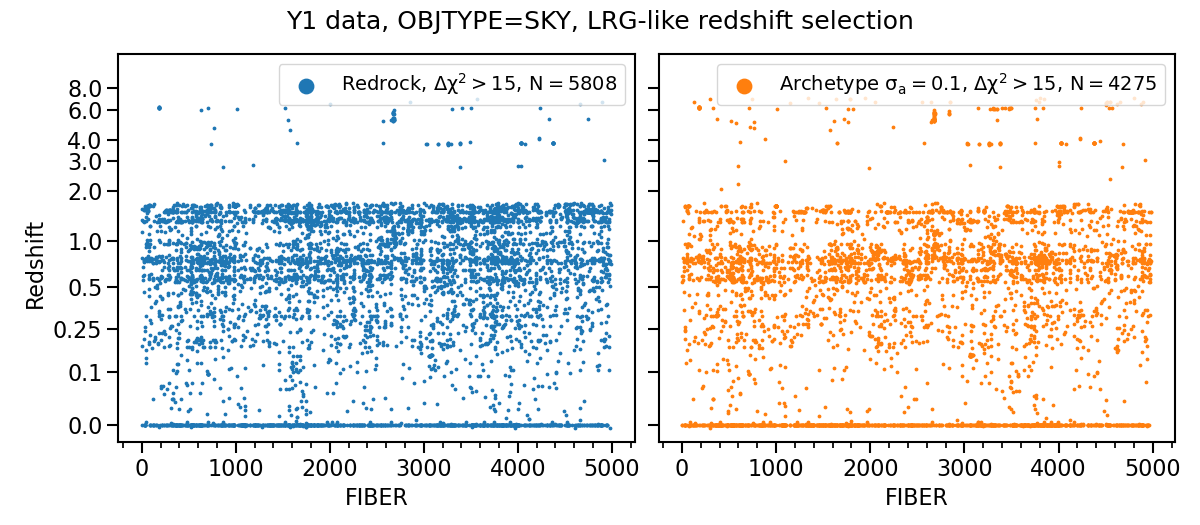}\\
 
    \caption{Redshift vs fibers for sky fibers (selected with \package{OBJSTYPE=SKY} criteria) for sky fibers from DESI Y1. In each panel, we also show the number of false positives selected using good redshift criteria (see Table~\ref{tab:z_success}) like true galaxy targets. The top panel shows results for \target{BGS}-like selection; the middle panel shows results for \target{ELG}-like selection, and the bottom panel presents the results for \target{LRG}-like redshift selection. The left panel (blue) shows the result for Redrock (without archetypes), and the right panels show the results of our archetype method (with a prior $\sigma_{\rm a}=0.1$ on polynomial coefficients, shown in orange) in all panels. \tb{It is evident that the archetype performs $6-40$\% better in rejecting false positive redshifts for sky fibers.}}
    \label{fig:skyfibers}
\end{figure*}

If the pipeline was perfect and the subtracted sky spectra were pure zero-mean uncorrelated noise, all templates would have the best solution of coefficients=0 and $\Delta\chi^{2}=0$. In this run, we first demonstrate our analysis by focusing exclusively on sky fibers to evaluate the efficacy of our methodology in the context of objects specifically targeted toward the blank sky. \tb{Note that the sky fibers are excluded from the final target redshift catalog. However, this is an important exercise to mitigate the false positive cases in our archetype method, as these spectra should not yield robust redshift. It also helps reduce false redshifts of real targets in sky-dominated wavelength regions. We also emphasize that some confident sky redshifts may be real serendipitous objects falling into sky fibers, though such occurrence is rare, and we do not focus on that here.} 

For this purpose, we use the same good redshift criteria (defined in Table~\ref{tab:z_success}) to understand the statistics of plausible redshift measurements for these sky fibers. Our expectation is to reduce the number of false positives for these fibers. 

We show the best-fit redshifts for these sky spectra as a function of fiber number in Figure~\ref{fig:skyfibers}. These best-fit redshifts point to the "attractor solutions" due to upstream pipeline imperfections, which risk pulling Redrock towards incorrect fits on very low S/N data. As elucidated in Table~\ref{tab:iron_skyfibers} and illustrated in Figure~\ref{fig:skyfibers}, our archetype approach yields a notably lower number of sky fibers with confident redshifts than the default PCA-only (without archetypes) run. This outcome aligns with expectations, as our method relies on physical galaxy models, and sky fibers typically do not correspond to any physical galaxy, being spectra of blank sky. In contrast, PCA templates exhibit a high degree of flexibility, allowing them to confidently fit such spectra, as evident by the data (see Redrock (without archetypes) column) in Table~\ref{tab:iron_skyfibers} and left panel of Figure~\ref{fig:skyfibers}. In the daily quality assessment, dedicated DESI collaborators look at these plots to understand the redshift performance. 

\begin{table}
\centering
  \caption{Redshift statistics of sky fibers from Y1 dataset}
  \addtolength{\tabcolsep}{-1.9mm}
  \begin{tabular}{cccccc}
    \hline
    Program&$N$&Selection&\multicolumn{3}{c}{$N (\rm false\, positives)$} \\
    &(skyfibers)&criteria&&(confident redshifts)&\\
    \hline
    & &&Redrock & Archetype  & Archetype \\
    &  &&&(no prior)&($\sigma_{\rm a} =0.1$)\\
    \multirow{2}{*}{dark}&$136,134$&\target{ELG}&2075&1980&1969\\
    &$136,134$&\target{LRG}&5808&4498&4275\\
    bright & 203,578&\target{BGS}&3184&1936&1902\\
    \hline
  \end{tabular}
  \label{tab:iron_skyfibers}
\end{table} 

In summary, our archetype method demonstrates a $6-40 \%$ improvement in mitigating erroneous redshift fits for sky fibers. This is a substantial improvement over the current redrock. For instance, when applying redshift success criteria akin to LRG on sky fibers (bottom panel Figure 10), \tb{the Redrock (without archetypes) run identifies $5808$ objects with robust redshifts. In contrast, our archetype method identifies only $4498$ without priors and $4275$ ($\sim 27$ percent improvement) with $\sigma_{\rm a} =0.1$ on polynomial coefficients.} This improvement in reducing false positive redshifts is consistently observed in our archetype approach (see Table~\ref{tab:iron_skyfibers} and right panel of Figure~\ref{fig:skyfibers}) with redshift selection akin to \target{BGS} (top right panel, $3184$ in archetype vs.~$1902$ in Redrock (without archetypes) and \target{ELG} (middle right panel, $1969$ in archetype vs. $2075$ in Redrock (without archetypes)) target classes. This outcome aligns with our rationale for utilizing priors—to reduce such misidentifications in the first place. 

Another improvement we observe with our archetype approach is that it reduces the accumulation of spurious redshifts around any particular redshift as a function of fiber number. In the bottom panel of Figure~\ref{fig:skyfibers}, we show results for sky fibers selected with redshift selection criteria akin to LRGs (see Table~\ref{tab:z_success}). We find that the Redrock (without archetypes) redshifts (bottom left panel) show spurious redshift accumulation between fibers $1400$ and $2000$. However, the archetype case has no such accumulation (Figure~\ref{fig:skyfibers}, bottom right panel). This can be attributed to the fact that archetypes represent physical galaxy spectra; consequently, they do not cluster around any fibers as the final fit can be any redshift with small $\Delta \chi^{2}$.

\subsubsection{Redshift success rate}\label{zsuccess}

Next, we estimate the redshift success rate for all DESI targets using the same redshift criterion defined in Table~\ref{tab:z_success} and compare our archetype results with success rates from the Redrock (without archetypes) run. We also note that we do not need to define a new criterion\footnote{Though this can be explored in the future when we run the archetype mode for all spectral types and redshifts.} to define the redshift success rate in our archetype approach while also improving the overall performance of the spectral fitting and redshift estimation pipeline. We present the comparison in Table~\ref{tab:iron_z_success}. We find that our archetype approach (with $\sigma_{\rm a}=0.1$) consistently yields a higher (by $0.1-0.8$\%) redshift success rate for almost all galaxy target classes. For example, the redshift success rate for \target{ELG\_LOP} is $70.90$ \% in the Redrock (without archetypes) run while it increases to $71.56$ \% in our archetype (with prior, $\sigma_{\rm a}=0.1$) approach ($\sim$ 0.7\% improvement). \modify{The lower success rate in the PCA-only model can, in part, be attributed to the misclassification of strong emission lines (e.g., see Figure~\ref{fig:spec2}). However, since the archetypes are derived from actual galaxy spectra, which include all the expected emission lines, can help find the correct redshift corresponding to their location in observed spectra.} \tb{Similar relatively higher redshift success rate is seen for \target{ELG\_VLO} and \target{ELG\_HIP}, the redshift success rate is higher in archetype mode.} Similarly, the success rate for LRGs is $98.81$\% in the archetype method compared to $98.79$\% in Redrock (without archetypes). \tb{Finally, for the QSO targets, the Redrock (without archetypes) delivers a $67.15\%$ redshift success rate, while it reduces to $66.84\%$ in archetype mode without prior. However, when we use prior $\sigma_{\rm a}=0.1$, the success rate increases to $67.11\%$.} This justifies our proposition that adding priors on Legendre polynomials in our model reduces quasar misclassification and does improve the overall performance for them (see section~\ref{priors}).

\modify{We also observe that the average redshift success rate is significantly lower for the \target{ELG\_LOP} class compared to other target classes. This discrepancy arises from a combination of the target selection algorithm and the redshift selection criterion. Firstly, ELG targets are selected using a combination of $grz$ colors, which serve as a proxy for the \oii flux and star formation. These targets extend nearly to the imaging depth and the color-based cuts are imperfect, leading to the inclusion of many faint targets with undetectable \oii flux for reliable redshift estimation.}

\modify{Additionally, the wavelength coverage of DESI targets poses another issue. For any target with $z>1.62$, the \oii line falls outside the spectrograph's range, resulting in no emission line being detected by DESI. Consequently, this results in redshift failure, as redshift efficiency is also dependent on the $S/N$ of the detected \oii flux.}

It is important to point out that in terms of relative percentage, these improvements are low; however, DESI will collect more than $\sim 30$ million galaxy spectra in 5 years, and even such marginal increase will translate to very large numbers in absolute terms. At the same time, the method also resolves the inherent issues with the current PCA-based method we discussed in section~\ref{archetype_model}. Furthermore, within our archetype approach, the redshift success rate is consistently higher for runs with prior (rightmost column of Table~\ref{tab:iron_z_success}) than without prior (middle column of Table~\ref{tab:iron_z_success}) on polynomial coefficients. This should not surprise us, as putting priors on the polynomial coefficients allows us to restrict their freedom, which reduces the unphysical modeling of input spectra. 

\begin{table}
\centering
  \caption{Test run on DESI targets from Y1 dataset}
  \begin{tabular}{ccccc}
    \hline
    Target&N&\multicolumn{3}{c}{Redshift Success Rate(\%)} \\
    \hline
    & &Redrock & Archetype  & Archetype \\
    & &&(no prior)&($\sigma_{\rm a} =0.1$)\\
    \target{BGS\_FAINT}&227,784&98.78&98.34&98.78\\
    \target{BGS\_BRIGHT}&562,848&98.64&98.53&98.72\\
    %\target{ELG}&384,234&73.64&74.21&74.21\\
    \target{ELG\_LOP}&340,064&70.90&71.55&71.56\\
    \target{ELG\_VLO}&44,170&94.63&94.68&94.68\\
    \target{ELG\_HIP}&94,046&75.76&76.29&76.29\\
    \target{LRG}&233,784&98.79&98.77&98.81\\
    \target{QSO}\footnote{Also see the footnote of Table~\ref{tab:z_success}.}&238,775&67.15&66.84&67.11\\
    
    \hline
  \end{tabular}
  \label{tab:iron_z_success}
\end{table}

\subsection{Further Algorithmic Tests}\label{more_tests}

We also explored various algorithmic adaptations by changing the model parameters and including additional complexities within our archetype-based galaxy redshift fitting method. The first test involved the variation of the number of best redshifts ($N_{\rm zbest}$, see Figure~\ref{fig:method_flowchart}) on which the archetype-based model should be run. We varied the $N_{\rm zbest}$ redshifts from $3$ to $15$. We observe that the redshift performance got worse for $N_{\rm zbest}\leq6$ compared to the Redrock (without archetypes). This outcome is likely attributable to the possibility that a small number of PCA-based best redshifts may not include the correct redshifts, particularly if the initial fit was unphysical. Next, increasing $N_{\rm zbest}$ to $9$ significantly improved the performance compared to Redrock (without archetypes) without compromising ($\sim 2$ times higher than default run) much on runtime, and we use this value for our final analysis. On the other hand, increasing $N_{\rm zbest}$ to $12$ and $15$ did not improve the $\chi^{2}$ values, and the best redshifts remained unchanged while significantly ($\sim 3$ times more than default run) increasing the runtime.   

Next, we also explored including additional nearest neighbors in our method to use \tb{multiple nearby archetypes in superposition so as to achieve a "sub-grid" interpolation. For that purpose, after getting the best archetype fit for a given spectrum (i.e., the archetype corresponding to the minimum $\chi^{2}$), we consider $N$-nearest archetype neighbors around that in $\chi^{2}$ space. Then, we construct a model that is a linear combination of these nearest archetypes and the Legendre polynomials and estimate the new $\chi^{2}$.} This is to test if increasing the number of archetypes improves the overall spectral fitting for galaxies further. However, we observe very slight improvements in $\chi^{2}$ in some cases, albeit at the expense of significantly higher runtime ($\sim 3$ times higher than a single archetype run), as it entails an extra computational step. The slight reduction in $\chi^{2}$ is expected, as adding more nearest neighbors introduces additional degrees of freedom into the equation. Consequently, considering the insignificant gain in best fit at the expense of increased runtime, we decided to run the algorithm with just one archetype. 

We also ran our method only with archetypes without Legendre polynomials, which performs worse than the current approach. This is expected, as archetypes alone can only model the shape and features of the spectrum and not the pipeline defects because it does not have the same flexibility as the Legendre polynomials in absorbing those defects accurately. 

Another test was to include higher-order (quadratic and above terms, see Eqn~\ref{eqn:per_camera_model} and Figure~\ref{fig:method_flowchart}) polynomial terms in the model. While it improves the $\chi^{2}$ in some cases and therefore, the best redshifts also change, however, it also yields some unphysical negative flux fittings as the combination of negative coefficients and higher degree polynomials can overfit the data. This also comes from the fact that the linear combination polynomial basis vectors can fit almost any curve. We plan to explore this in more detail using the NNLS-like method in the future.

\section{Discussion}\label{discussion}

This is one of the first comprehensive modifications in the redshift fitting algorithm of DESI spectral data and extensive testing on DESI data to quantitatively assess its performance. This effort is of particular significance in light of the substantial scale of DESI observations anticipated over the next $3-4$ years, addressing crucial issues inherent to the current algorithm.

\subsection{Improved Galaxy Modeling and Redshift Estimation with archetypes}

In Section~\ref{archetype_model}, we provide the details of our new archetype model coupled with a per-camera polynomial fitting (see Eqn~\ref{eqn:per_camera_model}) for modeling DESI spectra. Our method is significantly different from previous studies \citep{cool13, hutchinson2016} that have used archetypes to fit the low-resolution spectra and estimate their redshifts. They allow the archetype coefficients to take negative values, which is susceptible to yielding unphysical fits occasionally. They also do not include polynomial modeling of spectra in individual cameras and fail to accurately absorb the CCD and throughput offsets. Moreover, we also add priors (section~\ref{priors}) on our polynomial coefficients that help us reduce false positives in sky fibers and the misclassification of quasar spectra. Since the archetypes represent physical galaxy spectra with emission (such as \oii, \oiii, $H\alpha,\, H\beta$) and absorption features ($4000$~\AA\, break, Ca-K, H lines) characteristics to their physical properties, they seem to better at fitting those features in observed spectra and provides less flexibility in fitting unphysical features (see Figure~\ref{fig:spec2}). However, the reduced flexibility lacks the ability to capture some peculiar galaxy spectra that might also be detected serendipitously, which we will discuss in the next section. We ran our redshift fitter with the archetype-based approach on various small and large datasets and compared its performance against Redrock (without archetypes) as elucidated in Section~\ref{archetypefit}.

To quantify the improvement in redshift estimates with our archetype model, we performed a large test on DESI tiles across diverse datasets. Notably, these enhancements are achieved without changing the existing \textit{good redshift criteria}, though this can be further explored in the future. An extensive analysis of our method on tiles observed on multiple nights during the survey validation phase (see Section~\ref{repeat_observations}) shows that the new method yields fewer catastrophic failures ($10-30$ \%, see Table~\ref{tab:iron_z_success} and Figures~\ref{fig:bgs_bright_repeat},~\ref{fig:elg_repeat},~\ref{fig:lrg_repeat}) while simultaneously increasing the redshift purity for all galaxy subclasses, as it can deliver better models for the CCD discontinuities (see Figure~\ref{fig:spec1}). The other noteworthy improvement is in sky emission line regions. In cases where PCA-only (without archetypes) templates might erroneously fit the noise as an actual spectral feature, the archetype method inherently does not find physical features and, therefore, reduces false positive redshifts. Next, we also find that the new method is more effective ($10-20$\% in a relative sense) in reducing catastrophic redshift failure for visually inspected galaxy spectra (see Section~\ref{visual_tiles} and Table~\ref{tab:visual_comparison}).

Finally, when run on millions of DESI targets from the main survey (section~\ref{large_test}), our method shows an increased redshift success rate (see Table~\ref{tab:iron_z_success}) by $0.5-0.8$ \% for all galaxy target classes while yielding the similar success rate for QSOs as compared to the Redrock (without archetypes). Although the relative changes are small, the numbers are much larger in the absolute sense. The new method is relatively fast to process such a large number of targets within acceptable time-frames as described in section~\ref{code_details}. Furthermore, we find that adding a prior while solving for polynomial coefficients in our model further enhances the ability to reject false positive redshift measurements for sky fibers (i.e., the spectrum taken for blank sky regions, see Table~\ref{tab:iron_skyfibers}) while not reducing the redshift success rate for any galaxy subclass (sections~\ref{zsky} and ~\ref{zsuccess}). Based on our extensive tests, we conclude that the new archetype-based per-camera polynomial fitting method shows significant improvement over Redrock (without archetypes). Additionally, the method is generic enough that its functionalities can be easily extended to other similar large spectroscopic surveys in the near future. 

\modify{It is also important to compare the redshift success rate of the DESI pipeline with that of previous large surveys. For example, GAMA achieved approximately $97\%$ correct redshifts down to an apparent magnitude of 21 in the SDSS-$r$ band \citep{loveday2012}. Similarly, LRGs in the BOSS program had a redshift success rate of approximately $93\%$ for $i_{\rm fiber}<21.5$ \citep{bolton12}, while the DEEP2 survey delivered an overall success rate of $\geq 70\%$ \citep{cooper2006}. In comparison, the DESI pipeline, Redrock, consistently achieves higher success rates across all target classes. This improved performance is attributed to enhanced imaging, a refined target selection algorithm, robust PCA templates, better sky modeling, and also better resolution than BOSS.}

\subsection{Future Improvements and Extensions}\label{improvements}

It is also important to understand the shortcomings of our method. \tb{An important step would be to test our method independently of PCA-based template fitting. As described earlier, due to the runtime complexity, we still rely on a list of redshifts derived from the initial PCA-based redshift scan; therefore, if the true redshift is not present in that list, our method will fail. In the near future, we plan to optimize the \textit{BVLS} method and extend our method to all redshifts to perform an independent comparison with current DESI redshifts.} Another important improvement can be brought to the synthetic spectra dataset (Section~\ref{method}), which was generated using spectrophotometry from the datasets that were in place before DESI (like DEEP2 and AGES), but in the future, we will use DESI observations themselves. This may impact the overall quality and diversity of galaxies in these datasets as they are all flux-limited surveys and have restricted wavelength coverage and theoretical modeling of SEDs. This is an important point to remember while interpreting the current results.

With the ongoing DESI survey, we have collected millions of galaxy spectra with updated and more robust spectrophotometry and SED modeling. Notably, these spectra originate from the same instrument, facilitating the construction of a more consistent synthetic galaxy spectral dataset that aligns more accurately with observations. Concerted efforts are underway within the DESI collaboration to refine the galaxy templates based on the wealth of data accumulated from DESI. The initial value-added catalogs for these DESI galaxies, incorporating new state-of-art spectrophotometry (a.k.a \texttt{fastspecfit}\footnote{\url{https://fastspecfit.readthedocs.io/en/latest/}}, \citealt{moustakas2023}) and derived physical properties, are set to be released in near future \citep{moustakas2024}. This will be an important step in enhancing the accuracy and applicability of galaxy archetypes in redshift fitting. 

Another important issue is that our current ELG archetypes lack the broad diversity in line ratios associated with different metal transitions, which is usually associated with their stellar and gas properties. This may be critical for identifying and modeling different line ratios within the ELG subclass. Additionally, the current set of archetypes also lack AGNs and LINER-like galaxies and high redshifts ($z>1.5$) Lyman break (LBGs) galaxies or Lyman alpha emitters (LAEs). It is important to understand that the intricate theoretical modeling of such galaxies is extremely challenging; however, they are found in DESI datasets. Addressing this limitation represents another improvement that will be integrated into upcoming archetypes based on \texttt{fastspecfit} galaxy properties and line ratios in the future. Therefore, our archetypes may not comprehensively cover the vast diversity of galaxy spectra. Consequently, if we encounter galaxy spectra that are fairly different from our archetypes, we might miss them. \modify{However, running \textit{SetCoverPy} with appropriate weights while generating archetypes can be very useful for handling such outliers}. Even the default PCA-based templates were generated from a galaxy sample that excluded these special cases. However, the inherent flexibility of PCA allows adaptability in identifying diverse objects; it is not inherently designed for such specialized classification. Currently, a combination of visually inspecting the spectral features and redshift measurements from previous surveys is used to identify them. 

\modify{However, as described in section~\ref{zsuccess}, the overall low redshift success rate for ELGs is intrinsically related to their selection algorithm and stringent redshift criterion. The redshift criterion was developed to ensure a very low rate of catastrophic failures, which is crucial for clustering analysis. While relaxing this criterion might increase the success rate, it would also lead to an increased rate of catastrophic failures \citep{raichoor2023ELG}. To achieve significantly higher efficiency, the target selection algorithm must be refined, and the definition of reliable redshifts needs to be reconsidered.}

Our future strategy involves incorporating rare galaxy classes when generating PCA templates and galaxy archetypes. This step is deemed crucial for enhancing the precision of our redshift fitter, especially in light of upcoming surveys like DESI-II, which are being designed to target these peculiar galaxies \citep{ruhlmann-Kleider2024}.

In the current analysis, we are using only galaxy archetypes, though it is run on all redshifts regardless of spectral type. So, there still can be unphysical fitting for QSOs and stars, which have different spectral features than galaxy archetypes, and PCA-based $\chi^{2}$ can still be smaller. The challenge lies in constructing a set of archetypes that span the extensive diversity inherent to quasars. Currently, their unified model is not well understood and modeled \citep[see][for a recent review]{netzer15}. On the other hand, constructing archetypes for stars is also challenging, given the diversity of spectra from young to old stars. Another complexity is in modeling the spectra of cataclysmic variables (CVs) and white dwarfs (WDs), which show very distinct features from main sequence stars.

In the immediate future, a potential avenue for future enhancement would be redshift fitting in each amplifier within the DESI camera. Notably, the pipeline occasionally presents calibration challenges in amplifiers within the camera. This manifests again as an offset or gradient-like polynomial feature in regions within a given camera. This per-amplifier modeling is challenging due to wavelength overlap, and defining a method to split the spectra in each amplifier is not straightforward. This refinement can show further improvement in error absorption than what have gained now. Ongoing efforts are directed toward incorporating such intricacies in future releases of Redrock.

Additionally, we can explore the possible algorithms to construct archetypes from a given input set. Currently, we are using a generic classification algorithm (\textit{SetCoverPy}) for this, which may not be very efficient sometimes as the optimal reduced $\chi^{2}$ value for the distance matrix is not well defined. Therefore, developing an approach that includes clustering based on physical properties rather than relying solely on machine learning-based techniques can be more useful. However, this is quite challenging, given the difficulty in defining an N-dimensional parametric space that effectively splits galaxies or quasars based on their properties in that hyperplane. In any case, exploring this avenue is quite exciting and can offer potential insights into a more detailed understanding of galaxy and quasar classification. This also has implications for measuring their redshifts while also offering useful information on physical properties such as stellar mass or star formation rate.

Finally, as described earlier, our method is generic enough to quickly extend to other large surveys similar to DESI, such as WEAVE\footnote{William Herschel Telescope Enhanced Area Velocity Explorer.}, \modify{WAVES (planned in 4MOST survey), Prime Focus Spectrograph (PFS) on Subaru \citep{tamara2016} and MOONS\footnote{Multi-Object Optical and Near-infrared Spectrograph.} on Very Large Telescope (VLT, \citealt{cirasuolo2020}).} For instance, the upcoming WEAVE instrument \citep{dalton2014,jin2024} will have two cameras to obtain high-resolution spectra of galaxies and quasars. Given that our method can extract information on the number of cameras from the input spectrum, it can automatically fit the WEAVE spectrum in each of those two cameras and estimate the redshift and spectral class. We aim to explore this avenue and test its applicability to the WEAVE spectra in the near future.

\section{Summary of Conclusions}
In this paper, \tb{we have presented a modification to the existing redshift estimation algorithm for DESI spectra.} The \textit{archetype-based per-camera polynomial fitting} approach to fit the spectra of DESI galaxies and measure their redshifts and spectral classes. Our comprehensive tests demonstrate the efficacy of our method in mitigating unphysical model fits to galaxies and adeptly addressing pipeline defects introduced during the spectral reduction process. The main conclusions are as follows:

\begin{itemize}

\item The primary DESI redshift fitter Redrock occasionally suffers from the issue of yielding unphysical fits and inaccurate redshift and spectral type of galaxies. It also struggles to account for pipeline defects that are visible in the extracted spectra as vertical shifts or gradient-like polynomial features in the spectrum within one of the cameras due to CCD bias and camera throughput issues.

\item To develop more physical galaxy model fit and improve the redshift estimates, we employ an \textit{archetype}-based approach to model DESI galaxy targets. Furthermore, we introduce a \textit{per camera} polynomial spectral fitting to robustly absorb pipeline defects.

\item Our extensive tests show that the new method performs $10-40$\% better (\modify{for e.g., 31 vs. 19 in redrock vs archetype method for ELGs}) in reducing catastrophic redshift failure than the redrock (without archetypes) on tiles observed on several nights during the survey validation phase. Additionally, the redshift purity is consistently higher than the current redrock for these tiles in our archetype method. Furthermore, our method shows notable ability in finding precise redshifts for visually inspected high S/N coadded spectra of DESI targets.

\item \tb{We find that the archetype method performs marginally better on survey validation targets, which also confirms the overall robustness of the Redrock pipeline and, hence, the downstream DESI science results.}

\item Additionally, our method performs better on millions of DESI targets selected from the Y1 dataset. We observe an increased redshift success rate (by $0.5-0.8$ \%) for all galaxy target classes (\modify{compare $70.90\%$ vs $71.55\%$ for ELGs in redrock vs archetype method, respectively}) \tb{while maintaining similar success for QSO targets.} At the same time, we also find that our method reduces the number of false positive redshifts (by $5-40$\%) for sky fibers.

\item Finally, we explain the shortcomings and future improvements to our method and discuss the generic nature of our method and its applicability to other large upcoming surveys.

\end{itemize}

\section*{Data Availability}
 
The data plotted in the results is available at \url{https://doi.org/10.5281/zenodo.12007781}.

\section{Acknowledgement}
\modify{We thank the anonymous referee for providing insightful comments that have helped improve the clarity of the paper. All the computations were performed at the DOE's supercomputer facility NERSC, located at Berkeley Lab.}
This material is based upon work supported by the U.S. Department of Energy (DOE), Office of Science, Office of High-Energy Physics, under Contract No. DE–AC02–05CH11231, and by the National Energy Research Scientific Computing Center, a DOE Office of Science User Facility under the same contract. Additional support for DESI was provided by the U.S. National Science Foundation (NSF), Division of Astronomical Sciences under Contract No. AST-0950945 to the NSF’s National Optical-Infrared Astronomy Research Laboratory; the Science and Technology Facilities Council of the United Kingdom; the Gordon and Betty Moore Foundation; the Heising-Simons Foundation; the French Alternative Energies and Atomic Energy Commission (CEA); the National Council of Humanities, Science and Technology of Mexico (CONAHCYT); the Ministry of Science and Innovation of Spain (MICINN), and by the DESI Member Institutions: \url{https://www.desi.lbl.gov/collaborating-institutions}. Any opinions, findings, and conclusions or recommendations expressed in this material are those of the author(s) and do not necessarily reflect the views of the U. S. National Science Foundation, the U. S. Department of Energy, or any of the listed funding agencies.

The authors are honored to be permitted to conduct scientific research on Iolkam Du’ag (Kitt Peak), a mountain with particular significance to the Tohono O’odham Nation. 

\software{\texttt{Matplotlib}
  \citep{hunter07a}, \texttt{NumPy} \citep{harris20a}, \texttt{Scipy} \citep{virtanen20a}, \texttt{SetCoverPy} \citep{zhu2016}.}

\bibliographystyle{aasjournal}
\bibliography{main}

\appendix
\section{Archetypes vs Parent Galaxies Comparison in Color-color space}\label{archetype_vs_parent}

\begin{figure*}
    \centering
    \includegraphics[width=0.45\linewidth]{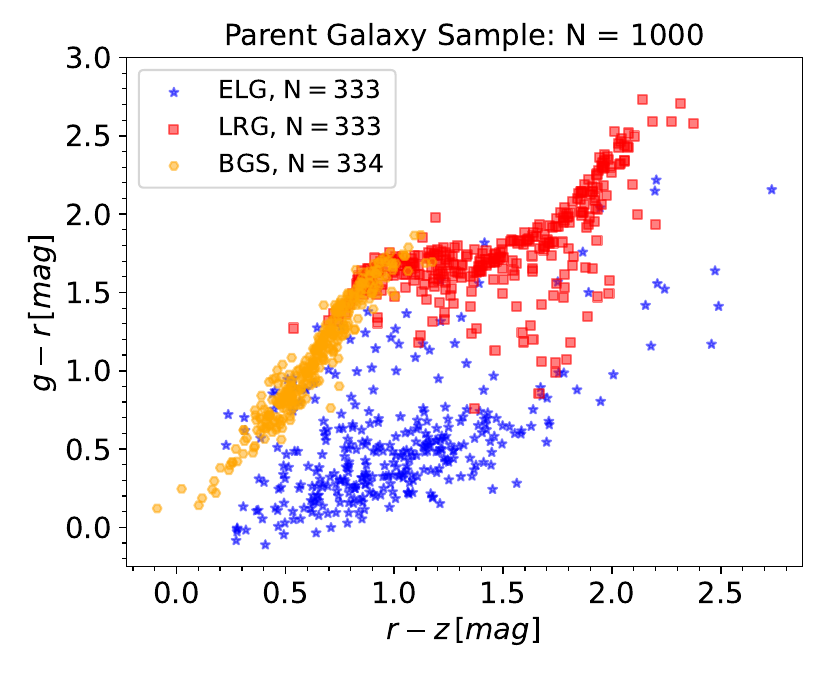}\includegraphics[width=0.45\linewidth]{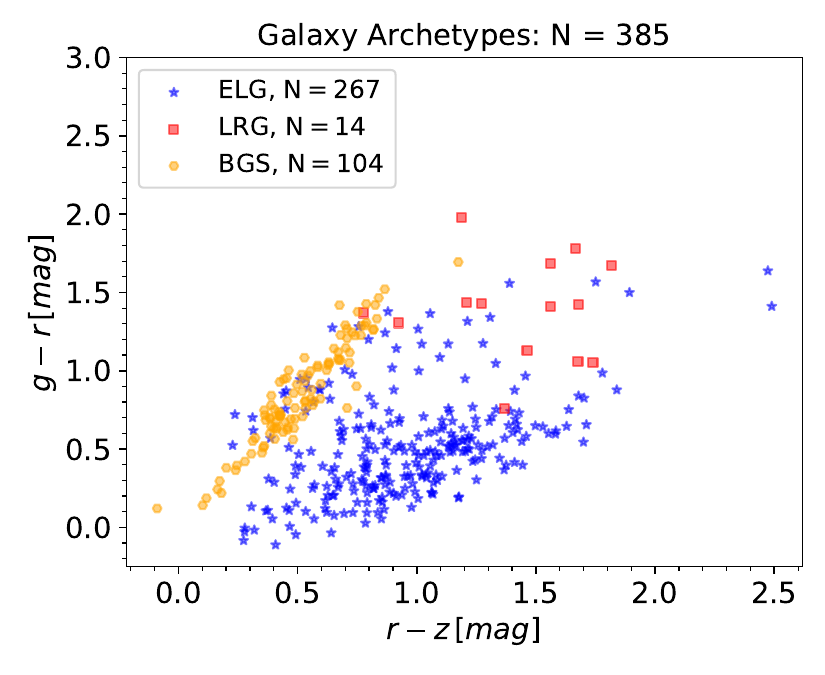}
    \caption{$g-r$ vs. $r-z$ color-color space comparison between parent (left) and archetype (right) galaxies. The archetypes obtained with \textit{SetCoverPy} span the properties of the parent sample well in this space. The colors are not corrected for interstellar extinction.}\label{fig:color_color_galaxies}
\end{figure*}

\begin{figure*}
    \centering
    \includegraphics[width=0.98\linewidth]{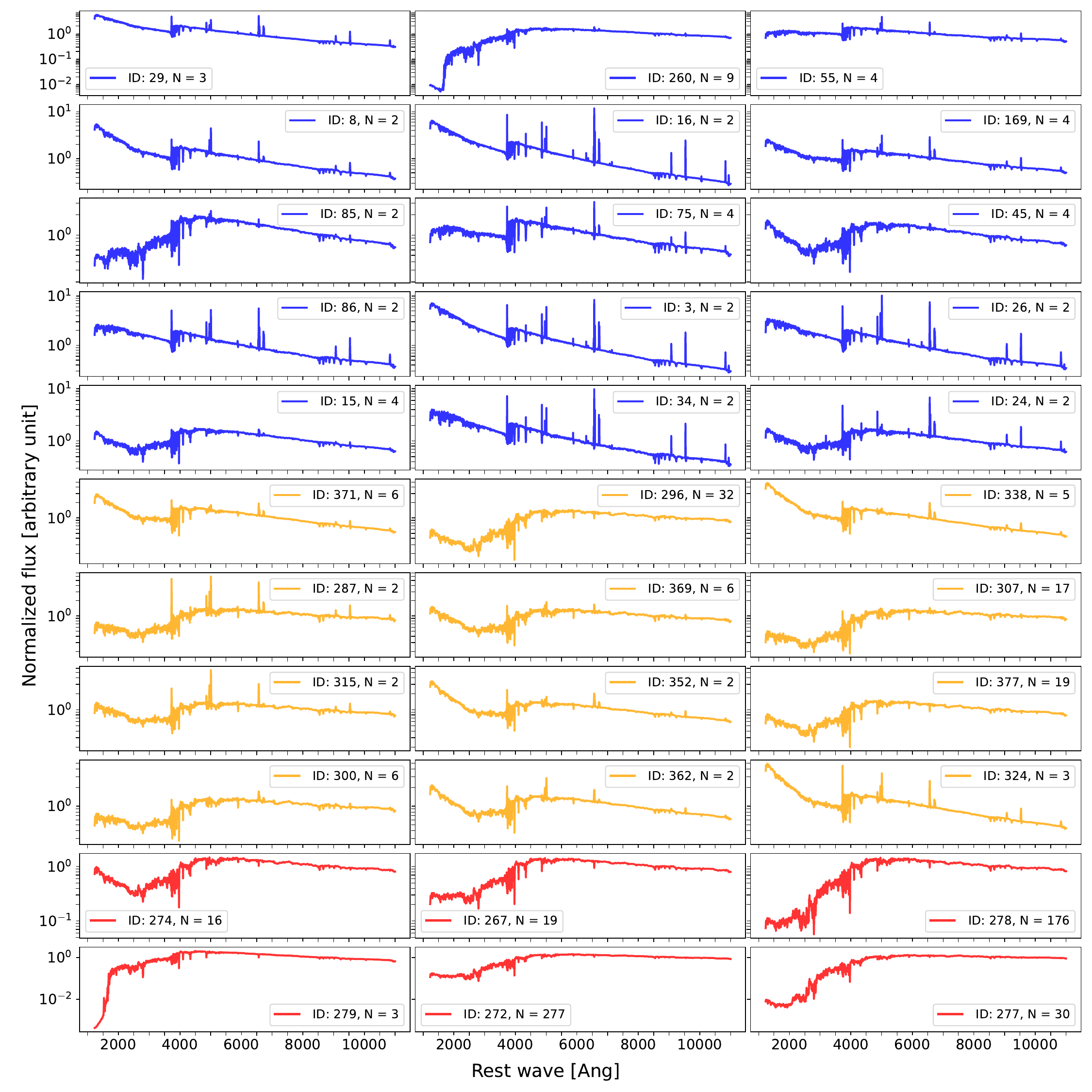}
    \caption{The rest-frame spectra of archetypes (shown as open circles in Figure~\ref{fig:archetypes}) that can represent more sources than themselves in the parent galaxy sample. The number $N$ shows how many galaxies in the parent sample the individual archetype can represent, i.e., with distances shorter than the minimum distances defined by $s^{2}$ threshold in \textit{SetCoverPy} method. We can see a larger diversity in ELGs than in LRGs.}
    \label{fig:example_archetypes}
\end{figure*}

To qualitatively assess the performance of \textit{SetCoverPy} in finding an optimal set of archetypes, we compared their distribution with the parent sample in physical space. We present one such comparison in Figure~\ref{fig:color_color_galaxies}, where we show the distribution of parent (left) and archetype (right) galaxies in $r-z$ vs. $g-r$ color-color space. The blue, orange, and red colors represent the ELGs, BGS, and LRGs, respectively. In the parent sample, we see that ELGs and LRGs occupy very different regions in the color-color space. A similar distribution is visible in archetypes, though we see a lot fewer LRGs in the archetype because their spectra usually do not show emission line features and continuum shape does not vary much. In fact, we see that there are no LRG archetypes with $g-r>2$; this is evident when we show below the spectra of a few LRG archetypes that span most of the parent LRGs. At the same time, the BGS sample lies somewhat in the \tb{intermediate} region in the color-color diagram. They are low redshift, very bright galaxies and can have galaxies ranging from high star-forming to passive galaxies. This observation underscores the reliability of \textit{SetCoverPy} in finding an optimally representative subset of galaxies. As pointed out in \citet{zhu2016}, since the fluxes are normalized, stellar mass vs. star formation rate may likely not be the most informative dimension to compare parent and archetype galaxy samples. Subsequently, we select $33$ random galaxies from our archetypes (shown in open circles in Figure~\ref{fig:archetypes}) and show their spectra in Figure~\ref{fig:example_archetypes}. One can see the diverse spectral features in ELGs and BGS, while LRGs do not show such diversity. Most of the LRGs are spanned by the three archetypes (ID: 272, 278, and 267). In each subpanel, we also show the number of spectra that the particular archetype can represent in the parent sample. We can see that a few LRGs can span almost all the LRGs in the parent sample. 

\section{Optimal prior value for polynomial coefficients in Archetype approach}\label{optimal_prior}

In order to select an optimal value for prior on coefficients of Legendre polynomials, we employ results of fibers targeting the blank sky. These sky spectra are used to construct robust sky models for the spectral reduction process. As illustrated in section~\ref{priors}, we ran our method on numerous tiles within the DESI Y1 dataset. This extensive analysis helped us understand the distribution of Legendre polynomials on hundreds of thousands of sky fibers. This comprehensive analysis enabled us to select an optimal value of the prior value for these polynomial coefficients. 

The sky fibers are expected only to have sky emissions and no emission lines associated with actual astronomical objects. Consequently, modeling spectra using Legendre polynomials and archetypes for these fibers should demonstrate minimal variability compared to the diversity observed in the spectra of galaxies. Therefore, the spread in the coefficient distribution allows us to constrain the flexibility of coefficients of the Legendre polynomial while fitting real spectra in the archetype method. As demonstrated in Section~\ref{priors}, the physical slope of the quasar continuum may occasionally be confused with slope features associated with pipeline defects due to excessive freedom in polynomial coefficients if no prior is used. The prior determined based on sky fibers restricts such freedom, thereby mitigating the misidentification of quasar spectra as galaxies or vice versa within the archetype approach.

\begin{figure*}
    \centering
	\includegraphics[width=0.95\linewidth]{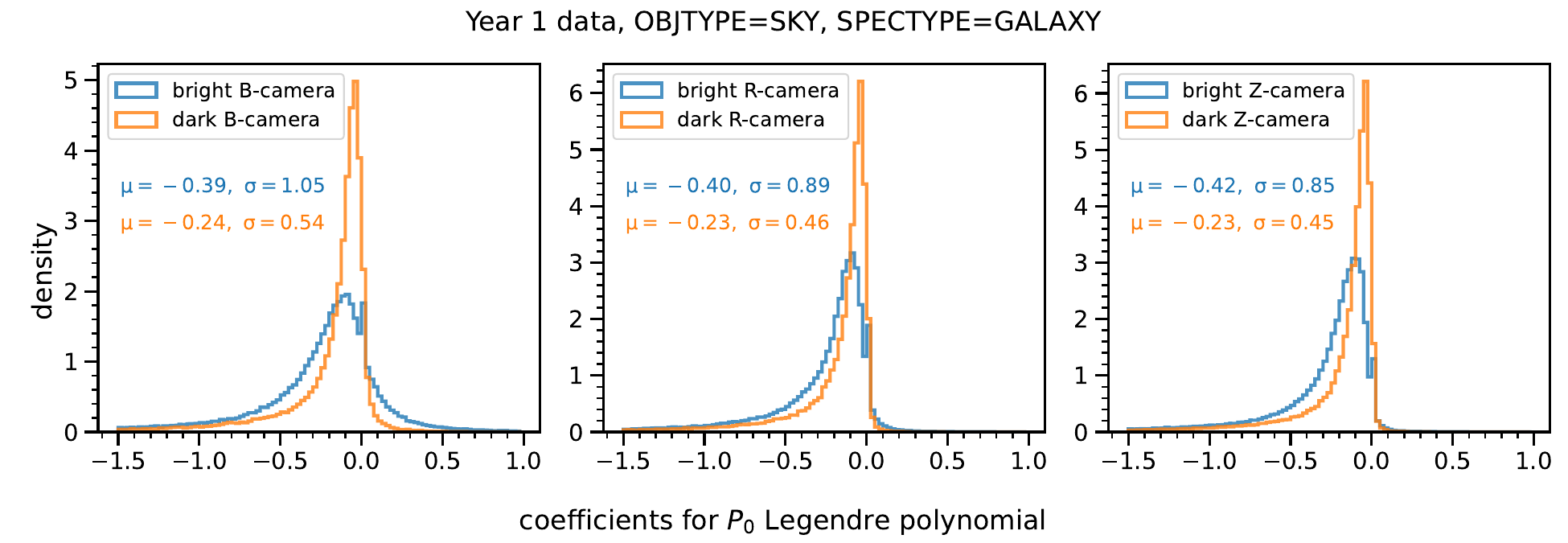}\\
	\includegraphics[width=0.95\linewidth]{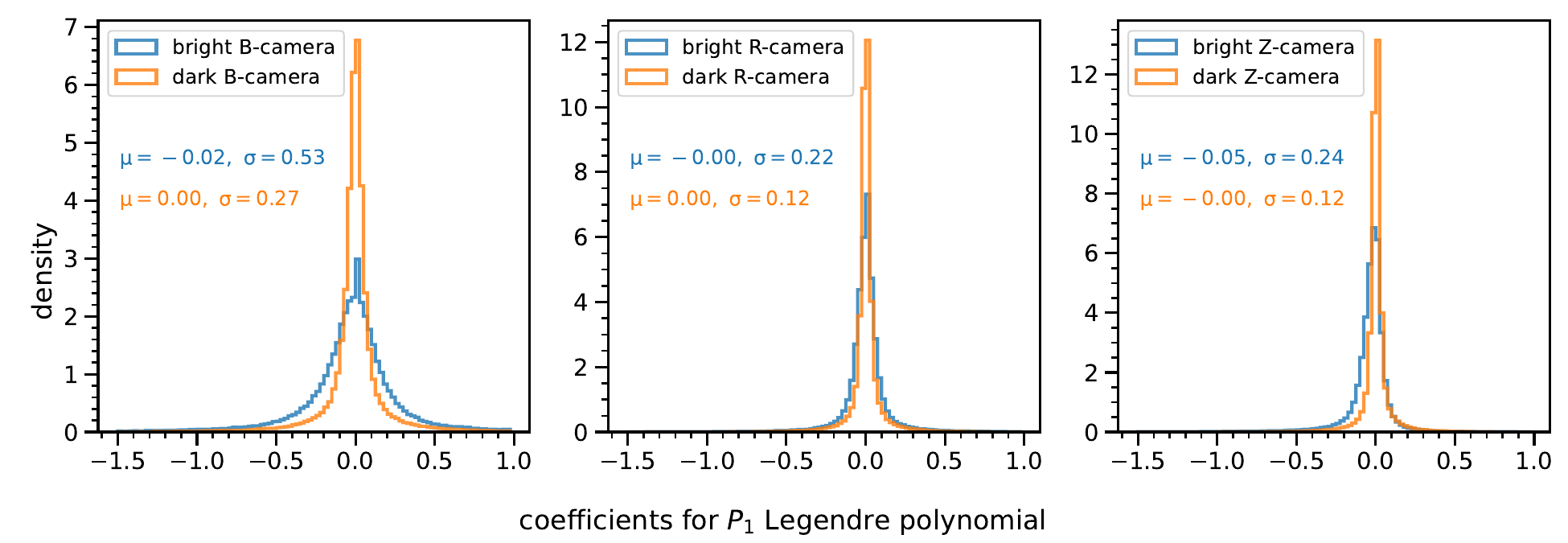}\\
    \caption{Distribution of Legendre polynomial coefficients (in each camera, constant term in the top panel and slope term in bottom panel) for sky fibers (for both dark and bright tiles) from the Y1 dataset. The spread of distribution varies between $0.1-0.5$ for dark tiles and $0.2-1$ for bright tiles. Based on this, we choose a prior value of $0.1$ (lowest) on polynomial coefficients for our current archetype method. More details are presented in section~\ref{priors}.}
    \label{fig:coeff_skyfibers}
\end{figure*}

We present the distribution of fitted coefficients of Legendre polynomials for these sky fibers in all three cameras in Figure~\ref{fig:coeff_skyfibers}. The observed distributions look approximately Gaussian, with a spread between $0.1$ and $1$. However, notable asymmetry is observed, potentially due to other pipeline defects introduced during the spectral reduction. Therefore, we have opted for a prior smaller ($\sigma_{\rm a}=0.1$) than these variances. This choice ensures that the Legendre coefficients take smaller values for the sky fibers as they are featureless spectra. \tb{In fact, we checked the distribution of Legendre coefficients for sky fibers after applying the priors, and we found that the average value is close to zero while the spread reduces to $\lesssim 0.05$. This is evidence that our regularization method (see also, section~\ref{priors}) has worked fairly well.}

\section{More performance metrics for SV phase data}\label{more_statistics}
\modify{The truth catalogs available for the spectra from the SV phase provide us the opportunity to define additional performance metrics such as precision, recall, and F1 scores for the galaxy datasets. We utilize these tables to compare the true redshifts with the redshifts measured by the pipeline. The redshift difference is calculated as $|\Delta v|=|\frac{z_{\rm visual}-z_{\rm pernight}}{1+z_{\rm visual}}|\cdot c$, where $z_{\rm visual}$ and $z_{\rm pernight}$ are the visual and measured redshifts of the targets observed on several nights during the SV phase (see Section~\ref{repeat_observations}). The \textit{good redshift criteria} is defined in Table~\ref{tab:z_success}. We use these two definitions to calculate the precision, recall, and F1 scores for all galaxy targets for both the Redrock (without archetypes) and archetype methods. We define the following quantities:}

\begin{align*}
\centering
    \text{True Positive, } \rm TP &= N\,(|\Delta v|<1000\, \rm km\,s^{-1}, \, X>X_{\rm o}) \\
    \text{False Positive, }\rm FP &= N\,(|\Delta v|>1000\, \rm km\,s^{-1}, \, X>X_{\rm o}) \, \text{; i.e., the catastrophic failures} \\
    \text{True Negative, } \rm TN &= N\,(|\Delta v|>1000\, \rm km\,s^{-1}, \, X<X_{\rm o}) \\
    \text{False Negative, } \rm FN &= N\,(|\Delta v|<1000\, \rm km\,s^{-1}, \, X<X_{\rm o}) \\
    \rm precision &= \rm \frac{TP}{TP+FP}\,; \,\,\,\, \rm recall = \frac{TP}{TP+FN} \\
    \rm F1\,score &= 2\cdot \rm \frac{precision \cdot recall}{precision +recall}
\end{align*}

\modify{where $X$ and $X_{\rm o}$ are defined in Table~\ref{tab:z_success}. We calculate these quantities for both methods and compile the values in Table~\ref{tab:recall_precison} for all galaxy targets. As one of our key objectives is to minimize the number of catastrophic failures (i.e., false positives), precision is more valuable than recall. Our results indicate that the overall precision of both the Redrock method (without archetypes) and the archetype method is quite high and very similar across all galaxy target classes. This finding again demonstrates that Redrock is already highly efficient, while our archetype-based model also achieves comparable high performance for DESI spectra, along with improved modeling of spectral defects.}

\begin{table*}
\centering
  \caption{Recall and Precision statistics for Redrock vs. Archetype for targets observed during SV phase}
  \begin{tabular}{ccccccccccc}
    \hline
    Target&\multicolumn{3}{c}{Redrock}&\multicolumn{3}{c}{Archetype}&\multicolumn{3}{c}{Archetype ($\sigma_{\rm a}=0.1$)} \\
    \hline
    & Precision&Recall & F1-score&Precision&Recall & F1-score&Precision&Recall & F1-score\\
    \target{BGS\_FAINT}&0.980&0.993&0.987&0.981&0.996&0.988&0.981&0.996&0.988\\
    \target{BGS\_BRIGHT}&0.977&0.995&0.986&0.977&0.996&0.987&0.977&0.996&0.986\\
    \target{LRG}&0.987&0.984&0.986&0.988&0.983&0.986&0.988&0.984&0.986\\
    \target{ELG\_LOP}&0.991&0.968&0.979&0.989&0.962&0.975&0.989&0.966&0.977\\
    \target{ELG\_VLO}&0.997&0.988&0.992&0.995&0.987&0.991&0.997&0.987&0.992\\
    \target{ELG\_HIP}&0.995&0.970&0.982&0.994&0.972&0.983&0.994&0.975&0.984\\
    \hline
  \end{tabular}
  \label{tab:recall_precison}
\end{table*}



\end{document}